\documentclass[aps,pre,showpacs,noshowkeys,amsmath,amssymb,amsfonts,superscriptaddress,longbibliography,reprint]{revtex4-1}
\usepackage[english]{babel}

\usepackage{graphicx}
\usepackage{bm}
\usepackage{physics}
\usepackage{mathtools}
\usepackage{gensymb}
\usepackage{array}
\newcolumntype{C}[1]{>{\centering\let\newline\\\arraybackslash\hspace{0pt}}m{#1}}
\DeclareMathOperator{\sgn}{sgn}
\setcitestyle{super}
\usepackage{caption}
\usepackage{subcaption}
\DeclareCaptionLabelSeparator{bar}{~\rule[-0.4ex]{0.2ex}{1em}~}
\DeclareCaptionLabelFormat{subfor}{\textbf{#2}}
\newcommand*\bfcaption[2]{\caption[#1]{\textbf{#1.}#2}}
\usepackage[dvipsnames]{xcolor}
\definecolor{UBcolor}{HTML}{007CC1}
\usepackage[colorlinks=true,pdfnewwindow=true,linkcolor=UBcolor,citecolor=UBcolor,urlcolor=UBcolor,breaklinks=true,linktocpage]{hyperref}
\usepackage[all]{hypcap}
\usepackage[nameinlink,capitalise]{cleveref}
\crefname{SI section}{SI Section}{SI Sections}
\Crefname{SI section}{SI Section}{SI Sections}
\begin{document}

\title{Scaling regimes of active turbulence with external dissipation}

\author{Berta Mart\'{\i}nez-Prat}
\altaffiliation{These authors contributed equally to this work}
\affiliation{Department of Materials Science and Physical Chemistry, Universitat de Barcelona, Barcelona, Catalonia, Spain}
\affiliation{Institute of Nanoscience and Nanotechnology (IN2UB), Universitat de Barcelona, Barcelona, Catalonia, Spain}

\author{Ricard Alert}
\altaffiliation{These authors contributed equally to this work}
\affiliation{Princeton Center for Theoretical Science, Princeton University, Princeton, NJ, USA}
\affiliation{Lewis-Sigler Institute for Integrative Genomics, Princeton University, Princeton, NJ, USA}

\author{Fanlong Meng}
\affiliation{Max Planck Institute for Dynamics and Self-Organization (MPIDS), D-37077 G\"ottingen, Germany}
\affiliation{CAS Key Laboratory for Theoretical Physics, Institute of Theoretical Physics, Chinese Academy of Sciences, Beijing 100190, China}

\author{Jordi Ign\'es-Mullol}
\affiliation{Department of Materials Science and Physical Chemistry, Universitat de Barcelona, Barcelona, Catalonia, Spain}
\affiliation{Institute of Nanoscience and Nanotechnology (IN2UB), Universitat de Barcelona, Barcelona, Catalonia, Spain}

\author{Jean-François Joanny}
\affiliation{ESPCI Paris, PSL University, Paris, France}
\affiliation{Laboratoire PhysicoChimie Curie, Institut Curie, PSL University, Sorbonne Universit\'es, UPMC, Paris, France}
\affiliation{Coll\`ege de France, Paris, France}

\author{Jaume Casademunt}
\altaffiliation{These authors contributed equally to the supervision of this work}
\affiliation{Departament de F\'{\i}sica de la Mat\`eria Condensada, Universitat de Barcelona, Barcelona, Catalonia, Spain.}
\affiliation{Universitat de Barcelona Institute of Complex Systems (UBICS), Universitat de Barcelona, Barcelona, Catalonia, Spain}

\author{Ramin Golestanian}
\altaffiliation{These authors contributed equally to the supervision of this work}
\affiliation{Max Planck Institute for Dynamics and Self-Organization (MPIDS), D-37077 G\"ottingen, Germany}
\affiliation{Rudolf Peierls Centre for Theoretical Physics, University of Oxford, Oxford OX1 3PU, United Kingdom}

\author{Francesc Sagu\'es}
\altaffiliation{These authors contributed equally to the supervision of this work}
\affiliation{Department of Materials Science and Physical Chemistry, Universitat de Barcelona, Barcelona, Catalonia, Spain}
\affiliation{Institute of Nanoscience and Nanotechnology (IN2UB), Universitat de Barcelona, Barcelona, Catalonia, Spain}

\begin{abstract}
	Active fluids exhibit complex turbulent-like flows at low Reynolds number. Recent work predicted that 2d active nematic turbulence follows universal scaling laws. However, experimentally testing these predictions is conditioned by the coupling to the 3d environment.  Here, we measure the spectrum of the kinetic energy, $E(q)$, in an active nematic film in contact with a passive oil layer. At small and intermediate scales, we find the scaling regimes $E(q)\sim q^{-4}$ and $E(q)\sim q^{-1}$, respectively, in agreement with the theoretical prediction for 2d active nematics. At large scales, however, we find a new scaling $E(q)\sim q$, which emerges when the dissipation is dominated by the 3d oil layer. In addition, we derive an explicit expression for the spectrum that spans all length scales, thus explaining and connecting the different scaling regimes. This allows us to fit the data and extract the length scale that controls the crossover to the new large-scale regime, which we tune by varying the oil viscosity. Overall, our work experimentally demonstrates the emergence of universal scaling laws in active turbulence, and it establishes how the spectrum is affected by external dissipation.
\end{abstract}

\date{\today}

\maketitle

Active fluids are able to flow spontaneously due to the internal stresses that are generated by their microscopic components \cite{Simha2002,Voituriez2005,Golestanian2011,Marchetti2013}. Despite being driven at the microscale, active fluids exhibit large-scale flows that often become chaotic \cite{Uchida2010,Thampi2014b,Doostmohammadi2018}. Such flows have been observed in a wide variety of systems including bacterial suspensions \cite{Dombrowski2004,Cisneros2007,Ishikawa2011,Wensink2012,Dunkel2013,Patteson2018,Li2019,Peng2020}, sperm \cite{Creppy2015}, mixtures of cytoskeletal components \cite{Sanchez2012,Henkin2014,Guillamat2017,Ellis2018,Lemma2019,Martinez-Prat2019,Tan2019,Duclos2020}, cell monolayers \cite{Doostmohammadi2015,Yang2016a,Blanch-Mercader2018}, and artificial self-propelled particles \cite{Nishiguchi2015,Kokot2017,Karani2019,Bourgoin2020}. All these systems operate at low Reynolds numbers, where inertia is negligible. Yet, by analogy to inertial turbulent flows in conventional fluids, this regime with active chaotic flows has been termed active turbulence. The comparison with the longstanding paradigm of inertial turbulence is compelling, and it motivated the search of emergent universal behaviour and scaling laws \cite{Kolmogorov1991,Frisch1995} in active turbulence.

These issues have been the subject of intense research and debate \cite{Giomi2015,Alert2020b,Bourgoin2020,Uchida2010,Wensink2012,Dunkel2013,Bratanov2015,Linkmann2019,Bardfalvy2019a,Thampi2013,Urzay2017,Carenza2020,Coelho2020,Krajnik2020}. Inspired by bacterial turbulence, initial work on extended Toner-Tu equations for polar flocks showed that these fluids exhibit flow spectra with scaling regimes. However, the corresponding scaling laws have non-universal exponents, whose values depend on model parameters \cite{Wensink2012,Grossmann2014,Bratanov2015,Slomka2017,James2018a,Linkmann2019,C.P.2020}. In contrast, recent theoretical work on active nematics predicted scaling laws with universal exponents, which are independent of the active fluid properties (e.g. viscosity or activity) \cite{Giomi2015,Alert2020b}.
Experimental evidence for such universal scaling, however, remained elusive, in part due to the challenge of accurately measuring flows for sufficiently broad ranges of scales.

In addition, most of the theoretical predictions and numerical studies have addressed the case of two-dimensional flows. As in classical turbulence, 2d systems are often embedded in a 3d setup. It is, therefore, necessary to separate the genuinely 2d features from the effects due to the coupling to the environment. In classical 2d turbulence, for instance, frictional dissipation with the environment not only cuts off the energy cascade towards large scales, but it also modifies the scaling exponent of the enstrophy cascade towards small scales\cite{Boffetta2005,Boffetta2012b}.

\begin{figure*}[tb]
\includegraphics[width=\textwidth]{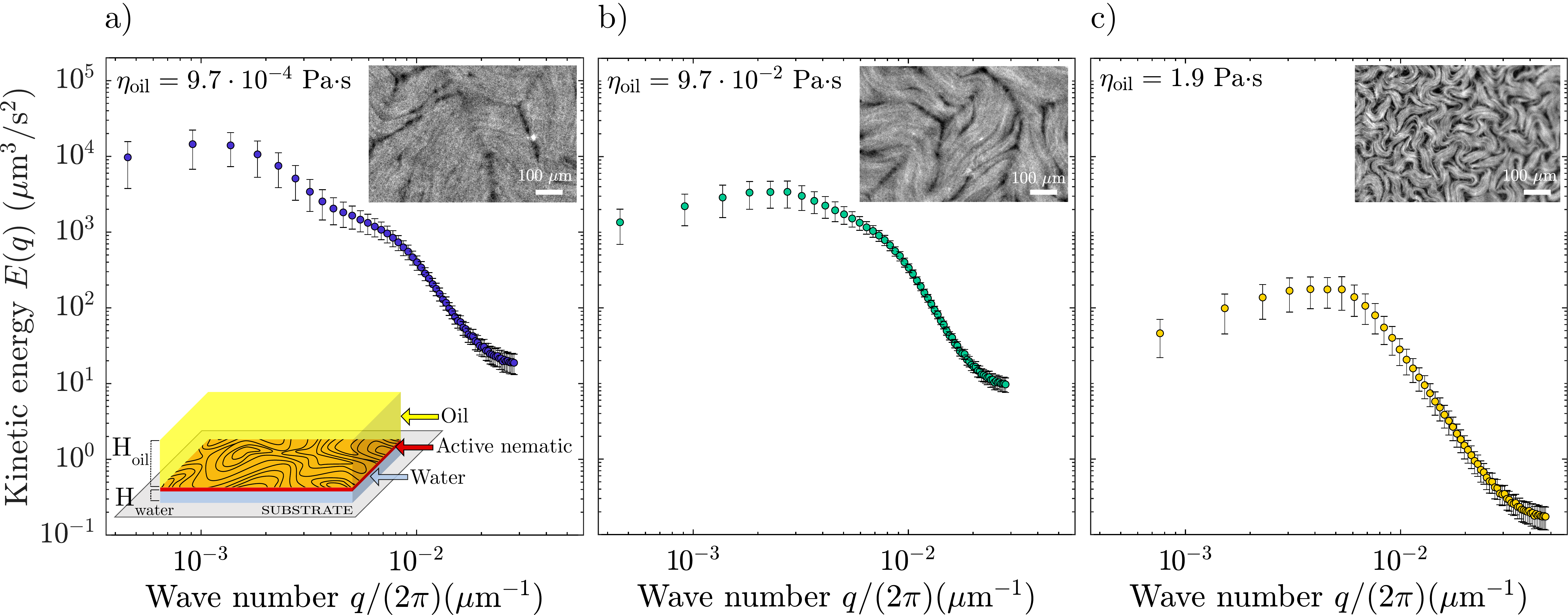}
  {\phantomsubcaption\label{Fig low-viscosity}}
  {\phantomsubcaption\label{Fig intermediate-viscosity}}
  {\phantomsubcaption\label{Fig high-viscosity}}
\caption{\textbf{Oil viscosity modifies the kinetic energy spectrum of active nematic turbulence}. \subref*{Fig low-viscosity}-\subref*{Fig high-viscosity}, Kinetic energy spectra of turbulent flows in an active nematic film in contact with an oil layer of low (\subref*{Fig low-viscosity}), intermediate (\subref*{Fig intermediate-viscosity}), and high (\subref*{Fig high-viscosity}) viscosity. Averages are over 500 frames. Error bars are s.d. In each panel, the top inset shows a representative microtubule fluorescence micrograph. The bottom inset in panel \subref*{Fig low-viscosity} shows a schematic of the experimental system. See also \hyperref[fig:movies1]{Movie S1}.
\label{Fig1}
}
\end{figure*}

Here, we address the influence of external dissipation on the scaling regimes in microtubule-based active nematics powered by motor proteins \cite{Sanchez2012,Martinez-Prat2019}. In this setup, the 2d active nematic is surrounded by layers of oil and water, two passive fluids. Our comprehensive experimental study spans a broad range of spatial scales and includes systematic variation of the oil viscosity. We also present a theoretical framework that incorporates the hydrodynamic coupling with the environment and provides an explicit expression for the full spectrum of the turbulent active flows. Altogether our results provide experimental evidence of universal scaling laws in 2d active turbulence, and determine the ranges where these  scaling laws are observable in terms of the physical parameters.

Our predictions include six different scaling regimes, which we classify in terms of three length scales: the average vortex size, the height of a external fluid layer, and a viscous length that controls whether dissipation is dominated by either the active or the external fluid. This analysis reveals that, in contrast to classic turbulence, external dissipation does not just introduce a small-$q$ cut off to the scaling behaviour, but it also yields a new scaling regime. In our experiments, we vary the oil viscosity over more than four orders of magnitude and observe three of the predicted scaling regimes. The remaining regimes might be observed either in alternative experimental setups or in other systems such as cell monolayers.

Beyond the scaling laws, the detailed comparison between theory and experimental data allows us to probe other open questions in active turbulence. Specifically, we seek to explain the emergence of a characteristic vortex size. The coupling to external fluids selects a characteristic wavelength at the onset of spontaneous flows \cite{Sarkar2015,Gao2015,Gao2017,Martinez-Prat2019}. However, we show that this selection mechanism cannot fully account for the average vortex size observed in the turbulent regime. Thus, our analysis suggests that nonlinear effects in the active fluid, possibly involving energy transfer across scales, also contribute to vortex size selection. Moreover, we establish the range of validity of our theory, as we observe that it correctly fits the data for an intermediate range of oil viscosities but fails for extremely low and high viscosities. Our measurements suggest directions for future improvements to the theory, such as including vortex-vortex correlations.

In our experiments, we prepare an active nematic film by self-assembly of micrometer-long stabilized microtubules at the interface between a $H_{\text{water}}\approx 40$ $\mu$m-thick water layer on a glass slide and a $H_{\text{oil}}\approx 3$ mm-thick oil layer open to the air \cite{Guillamat2016b} (bottom inset in \cref{Fig low-viscosity}, \hyperref[methods]{Methods}). The microtubules are bundled under the depleting action of polyethylene-glycol (PEG), which facilitates crosslinking by kinesin molecular motors clustered with streptavidin (\hyperref[methods]{Methods}). Fueled by adenosine triphosphate (ATP), the motors generate active shear stresses leading to extension and buckling of the microtubule bundles. The bundles then acquire local nematic order interrupted by half-integer topological defects, and the film exhibits disordered large-scale flows, which is the state of active nematic turbulence.

To study the statistical properties of active turbulent flows, we measure their so-called kinetic energy spectrum $E(q) \propto q \langle |\tilde{\bm{v}}(q)|^2\rangle$, where $\tilde{\bm{v}}(\bm{q})$ are the Fourier components of the flow field $\bm{v}(\bm{r})$, with wave vector $\bm{q}$ (\hyperref[methods]{Methods}). To explore the different flow regimes, we vary an external parameter, namely, the oil viscosity, while keeping the intrinsic active fluid parameters -- including motor, microtubule, and ATP concentrations -- fixed. Therefore, the so-called active length, which compares the active and elastic stresses, is kept constant in all of our experiments. Upon increase of the oil viscosity, the entire kinetic energy spectrum decreases (\cref{Fig1}), which is consistent with the previously-observed decrease of flow speed \cite{Guillamat2016b}. At low oil viscosities, the spectrum features at least three regimes: a large-scale (small-$q$) regime that is followed by a peak, an intermediate regime, and a crossover to a small-scale (large-$q$) regime (\cref{Fig low-viscosity}). As oil viscosity increases, the peak shifts to smaller scales, expanding the range of the large-scale regime and shrinking the intermediate regime until it can no longer be observed for high oil viscosities (\cref{Fig intermediate-viscosity,Fig high-viscosity}). In parallel to these changes in the flow properties, we also observe a higher density of defects for higher oil viscosities (top insets in \cref{Fig1}).

To understand our measured spectra, we develop a theoretical framework that accounts for the hydrodynamic coupling between the active film and the external water and oil layers, with their corresponding boundary conditions. Active flows in the nematic film induce flows in the passive layers which, in turn, influence the active-film flows (\cref{sketch}). Generalizing previous works \cite{Matas-Navarro2014,Guillamat2016b,Stone1995,Stone1995a,Lubensky1996}, we first obtain the Green function of the active flows subject to this feedback (\cref{Green's function}, see \cref{eq Green}). We then establish a relationship between the velocity power spectrum in our coupled three-fluid system and the vorticity power spectrum of an isolated active nematic film (\cref{spectrum derivation}, see \cref{eq velocity-vorticity-spectra}). The latter was previously predicted by Giomi using a mean-field theory approach based on decomposing the vorticity field into a superposition of $N$ uncorrelated vortices \cite{Giomi2015}. Based on simulation results, Giomi's theory assumes that each vortex has a uniform and size-independent vorticity $\omega_{\text{v}}$, and that vortex areas follow an exponential distribution with mean $a_* = \pi R_*^2$, where $R_*$ is the mean vortex radius.
For a configuration that has on average $N$ vortices over a total system area $\mathcal{A}$, we predict a kinetic energy spectrum (\cref{spectrum derivation})
\begin{equation} \label{eq energy-spectrum-main}
E(q) = \frac{B q R_*^4\, e^{-q^2 R_*^2/2} \left[ I_0\left(q^2 R_*^2/2\right) - I_1\left(q^2 R_*^2/2\right)\right]}{\left[q + \eta_{\text{oil}}/\eta_{\text{n}} \tanh(qH_{\text{oil}}) + \eta_{\text{water}}/\eta_{\text{n}} \coth(qH_{\text{water}})\right]^2},
\end{equation}
Here, $I_0$ and $I_1$ are modified Bessel functions, and $B = N \omega^2_{\text{v}}/(32\pi^3 \mathcal{A})$ is a prefactor related to the total enstrophy, which is independent of both the wave number $q$.

\begin{figure}[tb]
\includegraphics[width=\columnwidth]{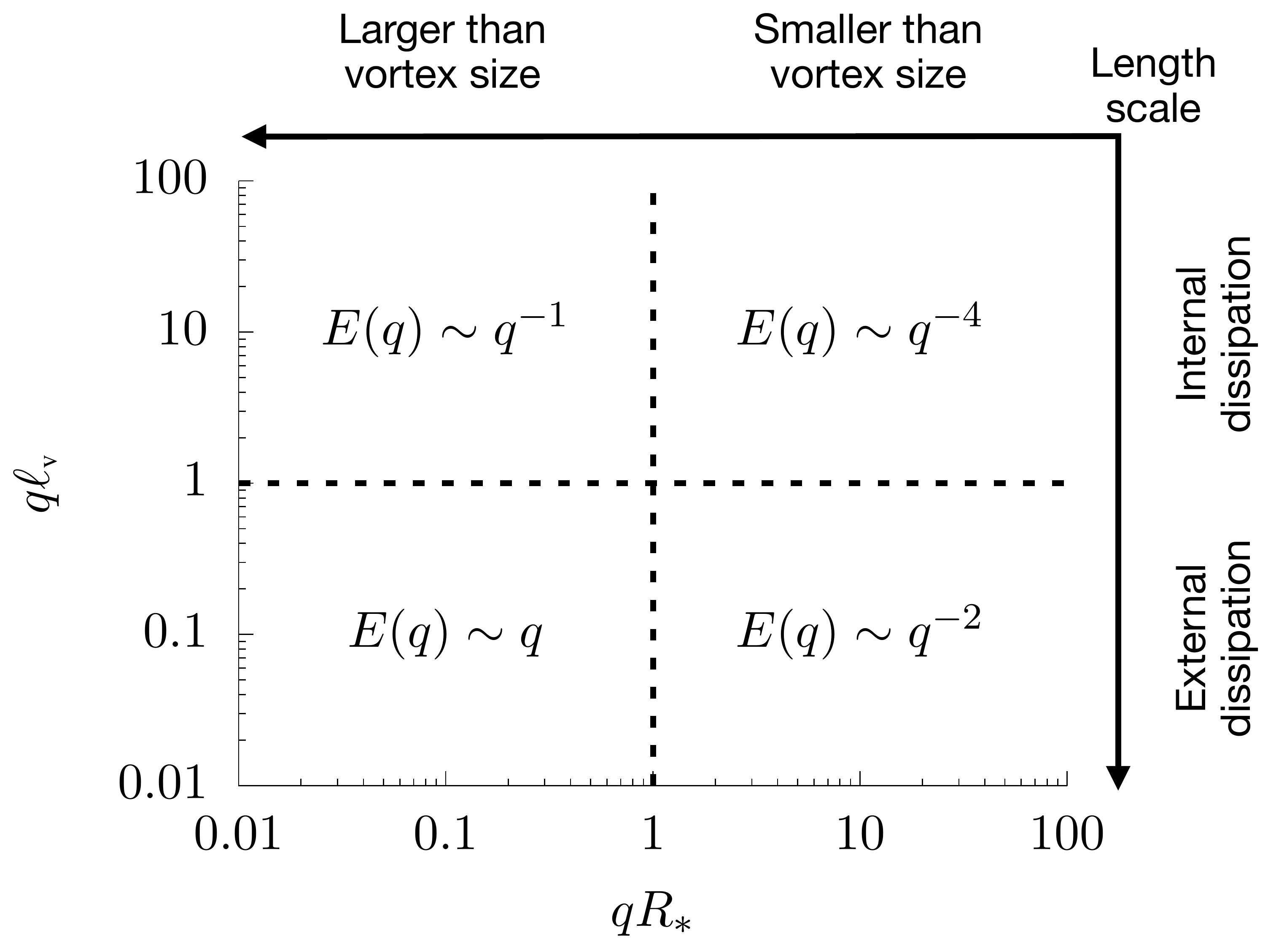}
\caption{\textbf{Scaling regimes of turbulent flows in an active nematic film in contact with an external fluid}. The different regimes are predicted at length scales ($\sim 1/q$) either larger or smaller than the mean vortex radius $R_*$, the viscous length $\ell_{\text{v}} = \eta_{\text{n}}/\eta_{\text{ext}}$, and the thickness $H$ of the external fluid layer. This figure summarizes the scalings in the thick-layer limit $q H\gg 1$; see \cref{Fig thin} for the predictions in the thin-layer limit $q H\ll 1$.
\label{Fig2}
}
\end{figure}

To discuss the scaling regimes predicted from \cref{eq energy-spectrum-main}, we consider a simplified situation with just one external fluid layer. The energy spectrum then depends on three length scales: the mean vortex radius $R_*$, the layer thickness $H$, and the viscous length $\ell_{\text{v}} = \eta_{\text{n}}/\eta_{\text{ext}}$ defined by the ratio of the two-dimensional viscosity of the nematic film, $\eta_{\text{n}}$, and the three-dimensional viscosity of the external fluid, $\eta_{\text{ext}}$. In the thick-layer limit, $q H\gg 1$, we summarize the predicted scaling regimes in \cref{Fig2}, which we discuss below. The thin-layer limit, $q H\ll 1$, is discussed in \cref{Fig thin} in \cref{scaling-laws}.

Active flows are characterized by different scaling laws at scales smaller and larger than the mean vortex size $R_*$ (horizontal axis in \cref{Fig2}). Furthermore, at length scales smaller than the viscous length $\ell_{\text{v}} = \eta_{\text{n}}/\eta_{\text{ext}}$, dissipation is dominated by the viscosity of the active film. As a result, in this regime, the scaling laws are those recently predicted and numerically demonstrated for isolated active nematic films \cite{Giomi2015,Alert2020b}, with no effect of the external fluid (top half in \cref{Fig2}). At length scales larger than $\ell_{\text{v}}$, however, dissipation in the external fluid dominates, yielding new scaling laws (bottom half in \cref{Fig2}).

\begin{figure*}[tb]
\includegraphics[width=0.75\textwidth]{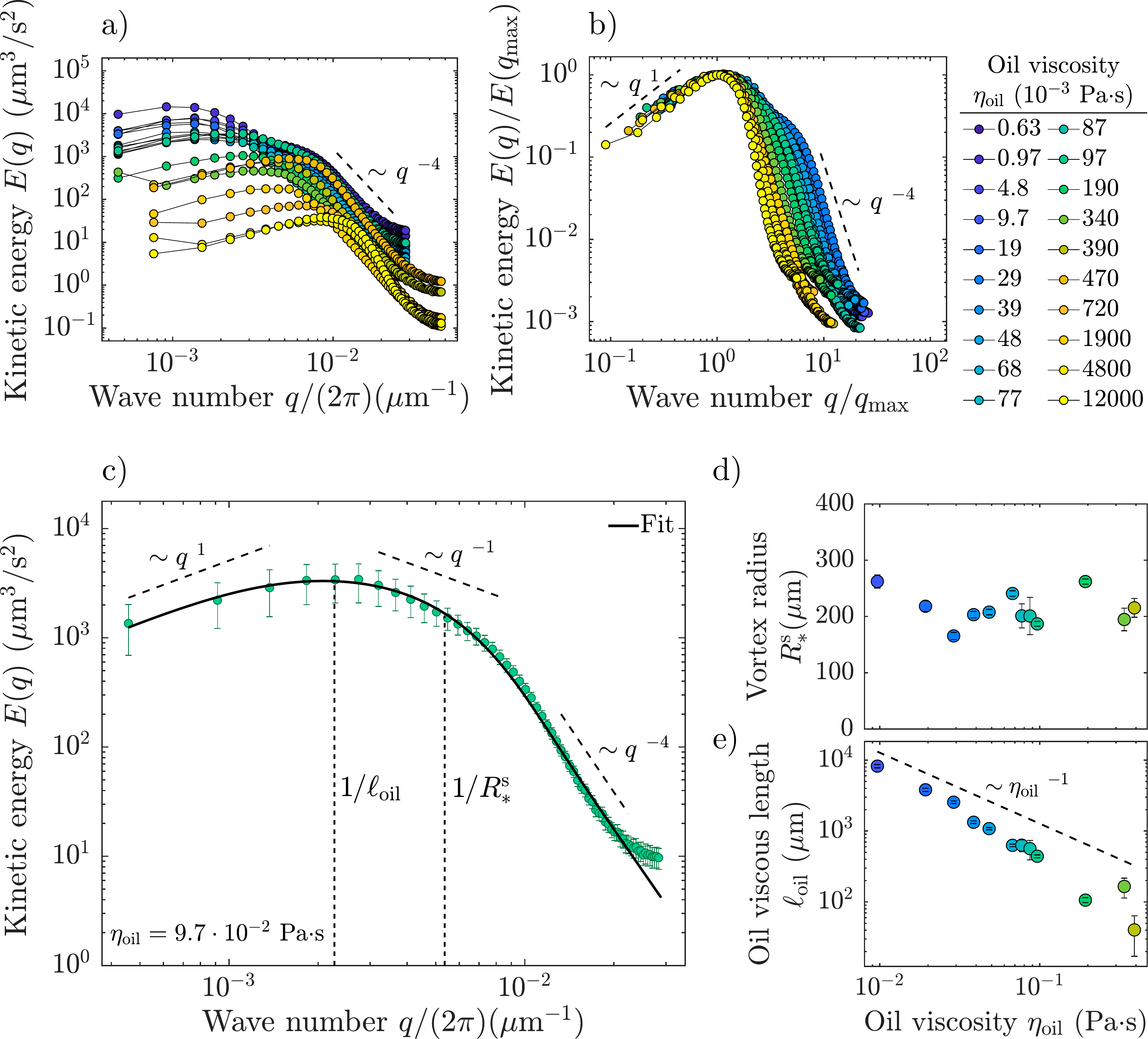}
  {\phantomsubcaption\label{Fig all-viscosities}}
  {\phantomsubcaption\label{Fig rescaled-spectra}}
  {\phantomsubcaption\label{Fig fit}}
  {\phantomsubcaption\label{Fig vortex-radius-spectra}}
  {\phantomsubcaption\label{Fig oil-length}}
\caption{\textbf{Oil viscosity tunes the scaling regimes of active nematic turbulence}. \subref*{Fig all-viscosities}, Kinetic energy spectra of turbulent flows in  an active nematic film in contact with a layer of oil, for 20 different oil viscosities. The data are averaged over 500 frames. \subref*{Fig rescaled-spectra}, Rescaling each spectrum by its maximum and its corresponding wave number clearly showcases the large-scale scaling regime. \subref*{Fig fit}, Fit of \cref{eq energy-spectrum-main} to a representative spectrum at intermediate oil viscosity (see fits for all oil viscosities in \cref{Fig spectra-all-viscosities}). As predicted by our theory (see \cref{Fig2}), the spectrum features signatures of three scaling regimes, separated by two crossover lengths: the mean vortex size $R_*^s$ and the viscous length $\ell_{\text{oil}} = \eta_{\text{n}}/\eta_{\text{oil}}$ (vertical dashed lines). Averages are over 500 frames. Error bars are s.d. \subref*{Fig vortex-radius-spectra},\subref*{Fig oil-length}, Mean vortex radius (\subref*{Fig vortex-radius-spectra}) and oil viscous length (\subref*{Fig oil-length}) obtained from the spectral fits in the range of intermediate oil viscosities in which the theory fits the data well. Error bars are s.e.m.
\label{Fig3}
}
\end{figure*}

In our experiments, the water viscous length $\ell_{\text{water}} = \eta_{\text{n}}/\eta_{\text{water}}$ is much larger than the largest length scale of our measurements: $q \ell_{\text{water}} \gg 1$ for all $q$ values. Therefore, the flows in the water layer do not produce any new scaling regimes in our experiments. In contrast, varying oil viscosity over orders of magnitude, we reach oil viscous lengths $\ell_{\text{oil}} = \eta_{\text{n}}/\eta_{\text{oil}}$ that fall within our measurement window. Consequently, the flows in the oil layer are responsible for some of the scaling regimes that we observe. Our measurements probe length scales that are smaller than the oil layer thickness $H_{\text{oil}}\approx 3$ mm. Thus, our experiments operate in the thick-layer limit, with scaling properties as predicted in \cref{Fig2}.

Consistent with these predictions, we experimentally observe $E(q)\sim q^{-4}$ at small scales, and $E(q) \sim q^1$ at large scales, for all oil viscosities (\cref{Fig all-viscosities,Fig rescaled-spectra}). For low and intermediate oil viscosities, we also observe signatures of the $E(q)\sim q^{-1}$ regime at intermediate length scales, larger than the vortex size $R_*$ but smaller than the viscous length $\ell_{\text{oil}}$ (\cref{Fig fit}). The $q^{-4}$ and $q^{-1}$ scaling behaviors are intrinsic properties of an active nematic film, respectively characterizing the small-scale flows inside vortices and the large-scale flows due to hydrodynamic interactions in the film \cite{Alert2020b}. In contrast, the $q^1$ scaling stems from the hydrodynamic coupling to an external fluid. All these scaling relations involve universal exponents, which are independent of the properties of the fluids. Consistently, by varying the oil viscosity, we tune the range of the different regimes without changing their scaling exponents.

Having demonstrated the scaling regimes, we quantitatively compare our prediction for the full energy spectrum (\cref{eq energy-spectrum-main}) to the experimental data. Knowing the values of $\eta_{\text{water}}$, $\eta_{\text{oil}}$, $H_{\text{water}}$, and $H_{\text{oil}}$, we use $B$, $\eta_{\text{n}}$, and $R_*$ in \cref{eq energy-spectrum-main} as fitting parameters. Despite its assumptions, our theory fits the data remarkably well for a wide range of intermediate oil viscosities ($9.7\times 10^{-3}$ Pa$\cdot$s $< \eta_{\text{oil}} < 0.39$ Pa$\cdot$s, see \cref{Fig spectra-all-viscosities}), as exemplified in \cref{Fig fit}. Visual inspection of the experiments suggests smaller vortices as $\eta_{\text{oil}}$ increases (see insets in \cref{Fig1} and \hyperref[fig:movies1]{Movie S1}). Yet, the mean vortex radius obtained from these fits, $R_*^{\text{s}}$, is independent of oil viscosity (\cref{Fig vortex-radius-spectra}). Finally, the oil viscous length $\ell_{\text{oil}} = \eta_{\text{n}}/\eta_{\text{oil}}$ decreases slightly faster than $\sim 1/\eta_{\text{oil}}$ (\cref{Fig oil-length}), indicating that $\eta_{\text{n}}$ weakly decreases with oil viscosity (\cref{Fig nematic-viscosity}). The physical origin of this dependence remains an open question.

To try to rationalize the fact that $R_*^{\text{s}}$ is independent of oil viscosity, we perform a linear stability analysis. Due to the coupling to the external fluid, the linear growth rate of the spontaneous-flow instability that powers active nematic turbulence acquires a maximum at finite wavelengths \cite{Sarkar2015,Gao2015,Gao2017} (\cref{selection,growth-rate}). This maximum selects a characteristic scale $\lambda_{\text{m}}$ of flow patterns at the onset of turbulence \cite{Martinez-Prat2019} (\cref{selected-wavelength}). This linear analysis predicts that $\lambda_{\text{m}}$ changes with oil viscosity, in contrast with experimental observations for $R_*$. However, nonlinear effects may modify this selected scale at later stages. In stationary fully-developed turbulence, earlier work \cite{Alert2020b} has shown that vortex size is given by the critical wavelength of the instability, $\lambda_c$. We now show that this length scale is largely independent of oil viscosity (\cref{selection,critical-wavelength}).

\begin{figure*}[tb]
\includegraphics[width=0.75\textwidth]{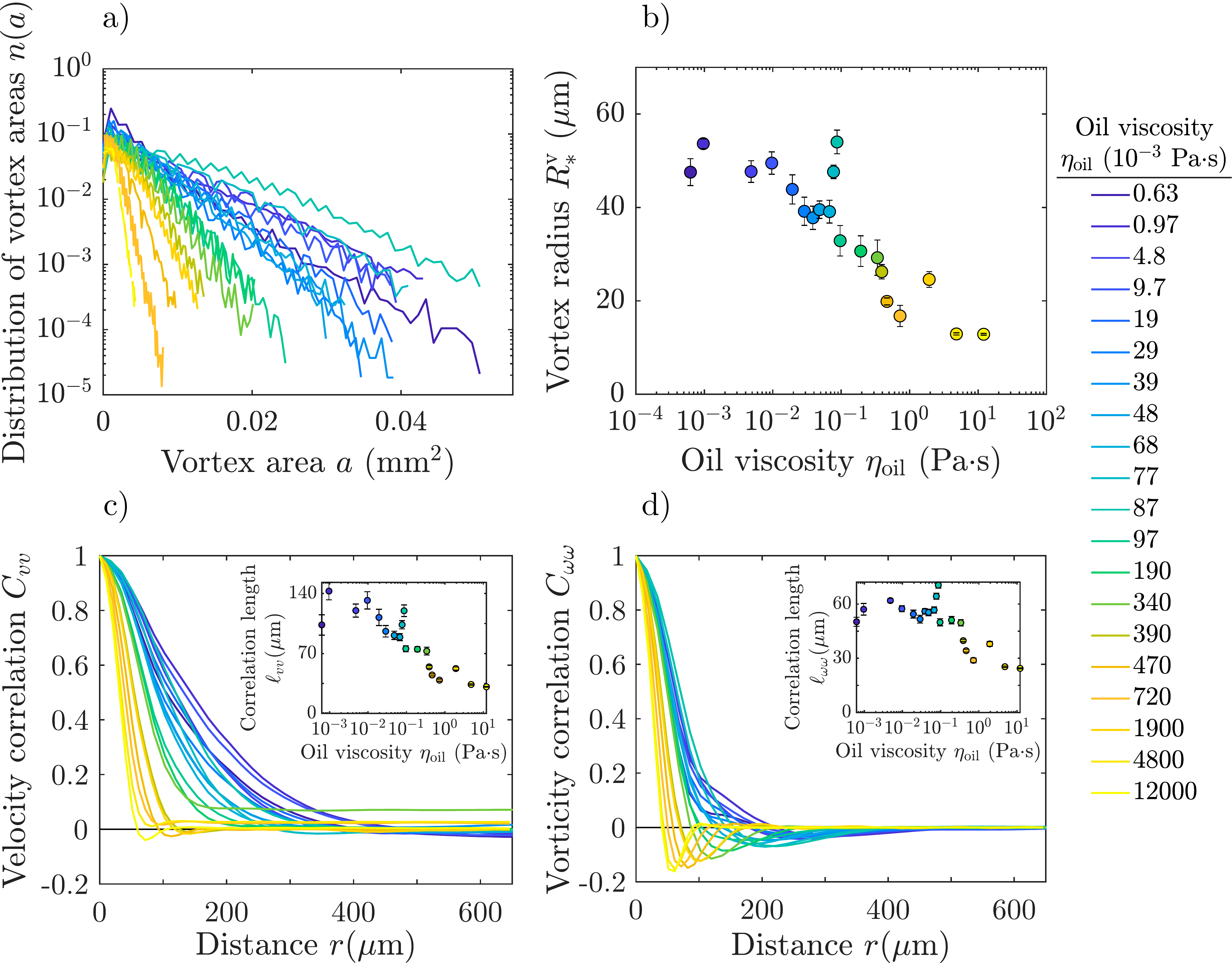}
  {\phantomsubcaption\label{Fig vortex-areas}}
  {\phantomsubcaption\label{Fig vortex-radius-distribution}}
  {\phantomsubcaption\label{Fig velocity-correlation}}
  {\phantomsubcaption\label{Fig vorticity-correlation}}
\caption{\textbf{Vortex size and correlation functions in active nematic turbulence}. \subref*{Fig vortex-areas}, Vortex area distributions in the active turbulence regime for 20 different oil viscosities. The distributions are obtained by measuring vortices in 500 frames. \subref*{Fig vortex-radius-distribution}, Mean vortex radius obtained from the exponential tails of the vortex area distributions in panel \subref*{Fig vortex-areas} (\hyperref[methods]{Methods}). Error bars are s.e.m. \subref*{Fig velocity-correlation},\subref*{Fig vorticity-correlation}, Spatial autocorrelation functions of the velocity (\subref*{Fig velocity-correlation}) and vorticity (\subref*{Fig vorticity-correlation}) fields, for all 20 oil viscosities. The data are averaged over 500 frames. The insets show the corresponding correlation lengths, defined by the conditions $C_{vv}(\ell_{vv}) = 0.5$ and $C_{\omega\omega}(\ell_{\omega\omega}) = 0.5$, respectively. Error bars are s.e.m.
}
\label{Fig4}
\end{figure*}

To address this issue experimentally, we directly measure the distributions of vortex areas $n(a)$ \cite{Guillamat2017,Lemma2019,Blanch-Mercader2018} (\cref{Fig vortex-areas}, \hyperref[methods]{Methods}). Fitting their exponential tails as $n(a) \propto \exp(-a/a^*)$ \cite{Giomi2015}, we obtain a mean vortex radius $R_*^{\text{v}} = \sqrt{a_*/\pi}$ as a function of oil viscosity (\cref{Fig vortex-radius-distribution}). The results show a decrease of vortex radius for high oil viscosities (\cref{Fig vortex-radius-distribution}), outside the range of validity of the presented model. For lower oil viscosities, within the range of validity of the model, the trend is weaker, suggesting that nonlinear effects contribute to vortex size selection in our system. Our analysis thus calls for further research to understand vortex size selection in active turbulence.

Finally, we use our measurements to examine some of the assumptions of the theoretical framework. We observe that vortex area distributions have exponential tails (\cref{Fig vortex-areas}), in agreement with both the theoretical assumption and previous experiments \cite{Guillamat2017,Lemma2019,Blanch-Mercader2018}. Here, we find that this feature does not change with the oil viscosity. We also measure the correlation functions for velocity and vorticity\cite{Thampi2013,Thampi2014,Hemingway2016} (\cref{Fig velocity-correlation,Fig vorticity-correlation}) and obtain the corresponding correlation lengths (insets in \cref{Fig velocity-correlation,Fig vorticity-correlation}), which exhibit dependencies on the oil viscosity that are very similar to that of the vortex size $R_*^{\text{v}}$ (\cref{Fig vortex-radius-distribution}). This observation suggests that these lengths are all proportional to one another (\cref{Fig lengths}), validating our theoretical assumption that there is a unique scale related to activity ($R_*$). However, at intermediate and high oil viscosities, we measure negative correlations of the vorticity field at distances comparable to, and even larger than, the vortex size (\cref{Fig vorticity-correlation}). Thus, our data calls for future work on the theory of active turbulence, going beyond mean-field approximations and accounting for vortex-vortex correlations. Other improvements to the theory could take into account the structure of defect cores, and include a more accurate treatment of flow alignment effects.

In summary, we have experimentally measured and theoretically explained  universal scaling laws in active turbulence, thus drawing parallels to classical turbulence. Specifically, we have found scaling regimes that are intrinsic to an active nematic film as well as other regimes that result from the coupling to an external fluid. In addition, we have developed a theoretical framework that provides the rationale for the different regimes and yields an explicit form for the turbulent spectra as a function of parameters. By fitting the predictions to the data, we have extracted the crossover length scales, and shown how they depend on the viscosity of the external fluid. Our analysis of these results paves the way toward addressing open questions in active turbulence, from vortex size selection to the role of vortex-vortex correlations, and to pursue a deeper understanding of the fundamental similarities and differences between inertial and active turbulence.

\section*{Acknowledgments}
The authors are indebted to the Brandeis University MRSEC Biosynthesis facility for providing the tubulin. We thank M. Pons, A. LeRoux, and G. Iruela (Universitat de Barcelona) for their assistance in the expression of motor proteins. B.M.-P., J.I.-M., and F.S. acknowledge funding from MINECO (projects FIS2016-78507-C2-1-P, AEI/FEDER, EU, and PID2019-108842GB-C22). J.C. acknowledges support from MINECO (projects FIS2016-78507-C2-2-P,  AEI/FEDER, EU, and PID2019-108842GB-C21) and Generalitat de Catalunya under project 2017-SGR-1061. B.M.-P. acknowledges funding from Generalitat de Catalunya through a FI-2018 PhD Fellowship. Brandeis University MRSEC Biosynthesis facility is supported by NSF MRSEC DMR-1420382. R.A. acknowledges support from the Human Frontier Science Program (LT000475/2018-C). F.M. and R.G. acknowledge support from the Max Planck Society. R.A. acknowledges discussions with the participants of the virtual "Active 20" KITP program, supported in part by the National Science Foundation under Grant No. NSF PHY-1748958.

\section*{Author contributions}
B.M.-P., J.I.-M., and F.S. designed the experiments. B.M.-P. performed the experiments. B.M-P., R.A., and F.M. analyzed data. R.A., F.M., R.G., J.-F.J., and J.C. developed the theory. All authors interpreted and discussed the results. R.A. and F.S. wrote the paper with input from all authors.

\section*{Competing interests}
The authors declare no competing interests.

\section*{Data availability}
All data are available from the authors upon request.

\section*{Code availability}
All codes are available from the authors upon request.

\bibliography{Bibl_Turbulence_experiments}
\onecolumngrid
\clearpage
\twocolumngrid

\section*{Methods} \label{methods}

\subsection*{Preparation of the active nematic}

Clusters of biotinylated kinesin-1 (K401-BCCP-6His) were assembled with tetrametric streptavidin at a $\sim$2:1 ratio. Afterwards, these molecular motors were mixed with adenosine triphosphate (ATP) and an {ATP-regenerating} system (Pyruvate Kinase/Lactic Dehydrogenase enzymes (PK/LDH) and phosphoenol pyruvate (PEP)). The non-adsorbing polymer polyethylene glycol (PEG) (20 kDa) was used as depleting agent. To avoid photobleaching and protein oxidation, a mixture of antioxidants (glucose, catalase, 1,4-dithiothreitol (DTT), Trolox and glucose oxidase) was also included. Biocompatibility of the active nematic layer with the oil interface was assured with the surfactant Pluronic F-127. This final suspension was mixed with $\sim$1.5 $\mu$m microtubules stabilized with guanosine-5'-[($\alpha$,$\beta$)-methyleno]triphosphate (GMPCPP) (Jena Bioscience, NU-405S), of which the 0.8\% was labeled with the Alexa-647 fluorophore. The scattered fluorescent microtubules formed a speckle pattern, from which we measure the velocity field through particle image velocimetry (PIV). Final compound concentrations are listed in \cref{tab:AN_composition}.

The active nematic (AN) was finally assembled by depositing a 1.5~$\mu$L droplet onto a polyacrylamide-functionalized glass slide within a 10 mm wide and 10 mm high polypropylene cylinder, and covered with 300~$\mu$L of polydimethylsiloxane oil (PDMS) with a viscosity in the range of 6.3$\cdot$10\textsuperscript{-4} Pa$\cdot$s to 12 Pa$\cdot$s. Micotubules spontaneously adsorbed onto the oil/water interface leading to the formation of the AN. The mean height of the water layer was obtained considering a cylinder and measuring the cross section area of the drop through fluorescence microscopy images. In the case of the oil layer, its height was directly measured with a millimetric ruler. We took the height at the center, where the AN drop is placed, neglecting the meniscus.

To have intermediate oil viscosities, oil mixtures were prepared. The final viscosity was estimated using the Arrhenius mixing rule ${\log(\eta_{12})=x_1 \log(\eta_1)+(1-x_1) \log(\eta_2)}$, where $\eta_{12}$, $\eta_{1}$ and $\eta_{2}$ are the oil viscosities of the oil mixture and of the mixed compounds $1$ and $2$, respectively, and $x_1$ is the molar fraction of oil $1$ \cite{Grunberg1949,Zhmud2011}.

\subsection*{Observation of the active nematic}

The active nematic layer was imaged by means of fluorescence microscopy (Nikon Eclipse Ti2-U) with an Andor Zyla 4.2 Plus camera controlled with the open-source software ImageJ Micro-Manager \cite{micro_manager_Edelstein2014}. As light source, we used a red LED coupled to a Cy5 cube filter. Images were acquired typically at a frame rate of 2 fps and with a spatial resolution of 2.14~$\mu$m/px. In the case of experiments with high oil viscosities ($>$1 Pa$\cdot$s) the frame rate was decreased to 1 fps and the spatial resolution was increased to 1.28~$\mu$m/px.

\begin{table}[tb]
	\begin{center}
		\begin{tabular}{|  C{3cm} | C{3cm}| }
			\hline
			\multicolumn{2}{|c|}{\textbf{Active nematic composition}} \\ \hline
			\textbf{Compound} & \textbf{Concentration} \\ \hline
			Streptavidin & 0.16 $\mathrm{\mu}$M \\
			Kinesin & 0.32	$\mathrm{\mu}$M \\
			DTT & 5.8	mM\\
			PEG (20 kDa) &	1.6	\% w/w\\
			PEP	& 27 mM\\
			Trolox &	2.1	mM\\
			MgCl\textsubscript{2} &	3.3	mM\\
			ATP &	1.5	mM\\
			PK/LDH &	27	IU/mL\\
			Pluronic & 0.44	\% (w/w)\\
			Glucose &	3.4	mg/mL\\
			Catalase &	0.040	mg/mL\\
			Glucose oxidase &	0.23	mg/mL\\
			Microtubules	& 1.3	mg/mL\\
			\hline
		\end{tabular}
	\end{center}
\caption{Composition of the active nematic.}
\label{tab:AN_composition}
\end{table}

\subsection*{Data analysis}
Raw images were treated with the open source software ImageJ \cite{ImageJ_Schindelin2012}. In general, light intensity was equalized by dividing the intensity of each frame by the time average of the sequence. Further noise removal was achieved with a mean filter with a width of 2--4~px. Afterwards, the velocity field was computed with PIVlab for Matlab \cite{PIV_Thielicke_2014}. PIV window size was set as 1/64\textsuperscript{th} of the lateral system size. To reduce noise, we used the option \textit{5 x repeated correlation}. The velocity field was finally smoothed using the \textit{smoothn} function developed by Garcia \cite{smooth_Damien2010}. The angle-averaged spectrum of the kinetic energy density per unit mass, $E(q)$, was computed from the obtained velocity fields as:
\begin{equation}
E(q)=L^2\pi q\langle|\tilde{\bm{v}}(q)|^2\rangle
\label{discrete},
\end{equation}
with
\begin{equation}
|\tilde{\bm{v}}(q)|^2=\frac{1}{2\pi}\sum_{\varphi}|\tilde{\bm{v}}(q,\varphi)|^2\Delta\varphi,
\end{equation}
where $L^2$ is the area of the field of view, $\langle \cdot \rangle$ indicates a time average, $\tilde{\bm{v}}$ are the Fourier components of the velocity $\bm{v}$, and $q$ and $\varphi$ are, respectively, the magnitude and azimuth of the wave vector $\bm{q}$. Note that the experimental Fourier transform is a discrete Fourier transform, whereas we use a continuous Fourier transform in the theory. This explains the difference in the prefactors between \cref{discrete} and \cref{eq kinetic-energy-spectrum-definition} in the Supplementary Information. For each experiment, we averaged a total of 500 frames.

Afterwards, \cref{eq energy-spectrum-main} was fitted to the experimental $E(q)$ with Mathematica v10.

Velocity and vorticity correlation functions ($C_{vv}$ and $C_{\omega\omega}$, respectively) were computed using the Wiener-Khinchin theorem: $C_{xx}(\bf{r})=\mathcal{F}^{-1}[|\tilde{x}(\bm{q})|^2]$, where $\mathcal{F}^{-1}[\cdot]$ denotes the inverse Fourier transform operator, and $\tilde{x}$ denotes either $\bm{v}$ or $\omega$.

Identification and characterization of vortices were carried out using the Okubo-Weiss (OW) parameter, as described previously\cite{Giomi2015,Lemma2019,Guillamat2017}. Briefly, OW was calculated as ${\text{OW}=(\partial_xv_x+\partial_yv_y)^2-4(\partial_xv_x)(\partial_yv_y)+4(\partial_xv_y)(\partial_yv_x)}$. Regions with $\text{OW}<0$ were vortex candidates. To determine whether such regions were vortices, we checked if they featured a singularity by computing the winding number $W(r)=1/2\pi\oint_0^{2\pi}\atan(v_y(r,\varphi)/v_x(r,\varphi))d\varphi$ of the velocity field at a distance $r=24~\text{px}$ from their center. If a region with $\text{OW}<0$ had a winding number $W\in[0.95,1.05]$, it was accepted as a vortex. The total area of each swirl was determined by the connected area with $\text{OW}<0$. This vortex locator algorithm, allowed us to extract a distribution of vortex areas $n(a)=N(a)/\sum_aN(a)$, where $N(a)$ is the number of vortices with area $a$. Since $n(a)$ follows an exponential distribution, we extracted a characteristic vortex area $a_*$ and radius $R_*^\text{v}=\sqrt{{a}_*/\pi}$.

\onecolumngrid 
\clearpage
\onecolumngrid

\begin{center}
\textbf{\large Supporting Information for}\\
\smallskip
\textbf{\large ``Scaling regimes of active turbulence with external dissipation’’}
\bigskip
\end{center}

\setcounter{equation}{0}
\setcounter{figure}{0}
\renewcommand{\theequation}{S\arabic{equation}}
\renewcommand{\thefigure}{S\arabic{figure}}

\twocolumngrid

\section*{Supplementary Text} \label{theory}

\subsection{Hydrodynamic Green's function}
\label[SI section]{Green's function}

\begin{figure}[tb]
\begin{center}
    \includegraphics[width=\columnwidth]{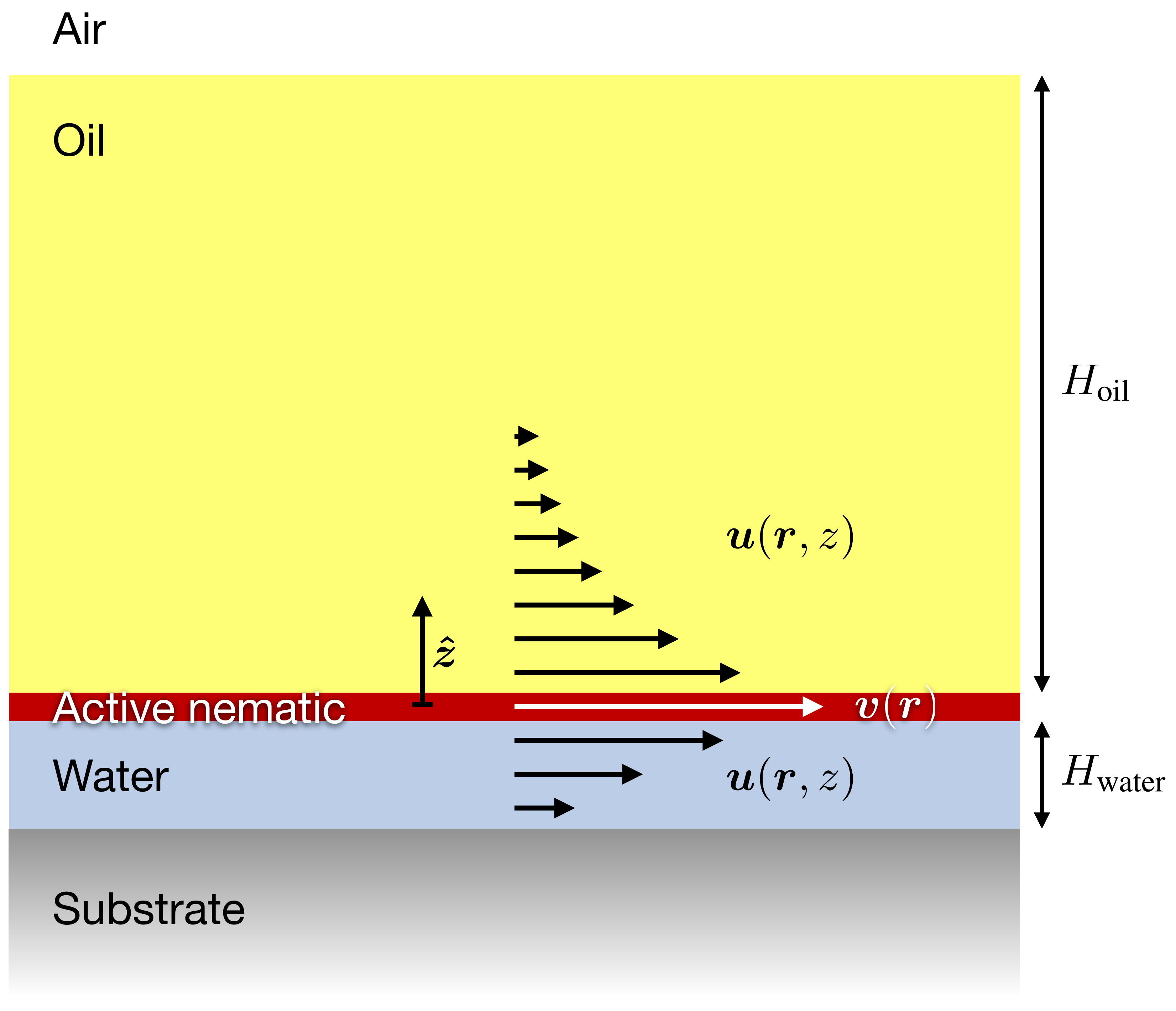}
\end{center}
\bfcaption{Schematic of the experimental setup and flow fields (side view)}{ The thicknesses of the fluid layers are not to scale. The actual thickness of the active nematic film is $h\approx 2$ $\mu$m, whereas the thicknesses of the passive fluid layers are $H_{\text{water}} \approx 40$ $\mu$m and $H_{\text{oil}} \approx 3$ mm. We treat the active nematic as a two-dimensional film. White and black arrows represent the flow fields in the active nematic film and in the passive fluid layers, respectively. The flow is planar and it penetrates into the oil and water layers. Here, we represented flow penetration according to \cref{eq layers-flow} for a planar wave number $q/(2\pi) = 5\cdot 10^{-3}$ $\mu$m$^{-1}$, which lies in the range of our experimental measurements.}
\label{sketch}
\end{figure}

In our experimental setup, a thin film of the active nematic fluid is in contact with a thicker layer of water underneath, and a much thicker layer of oil above. The water layer is supported by a solid substrate, and the oil layer is in contact with air (\cref{sketch}). Active flows in the nematic film induce flows in both passive fluid layers, which in turn, influence the flows in the active film. To account for this hydrodynamic coupling, in this section we obtain the Green function for the flow field in a two-dimensional viscous fluid film in contact with three-dimensional layers of other viscous fluids. This calculation generalizes previous work \cite{Lubensky1996} by considering the simultaneous hydrodynamic coupling to two fluid layers with different boundary conditions, as in our experimental setup (\cref{sketch}).

All flows in our system take place at very low Reynolds numbers. Therefore, inertial forces are negligible (Stokes limit), and momentum conservation reduces to force balance. For the active nematic film, with two-dimensional shear viscosity $\eta_{\text{n}}$ and incompressible flow field $\bm{v}(\bm{r})$, force balance can be written as
\begin{equation} \label{eq nematic-Stokes}
\eta_{\text{n}} \nabla^2 \bm{v} - \bm{\nabla}P + \bm{f}_{\text{water}} + \bm{f}_{\text{oil}} + \bm{f} = \bm{0}.
\end{equation}
Here, the first term accounts for the viscous stress within the nematic film, $P$ is the film's two-dimensional pressure, and $\bm{f}_{\text{water}}$ and $\bm{f}_{\text{oil}}$ are the viscous force densities exerted by the water and oil layers, respectively, on the active fluid film. Finally, $\bm{f}$ is the remaining force surface density acting on the film. In our system, this force density results from stresses associated with the orientational order of the nematic film, including elastic, flow-alignment, and active stresses. Here, we derive the Green function treating $\bm{f}$ as a source, without specifying its expression. Rather, we express the flow field of the nematic film in terms of the source force $\bm{f}$ as
\begin{equation}
v_\alpha (\bm{r}) = \int G_{\alpha\beta}(\bm{r}-\bm{r}') f_\beta(\bm{r}') \dd\bm{r}',
\end{equation}
where $G_{\alpha\beta}(\bm{r}-\bm{r}')$ is a hydrodynamic Green function, namely a generalization of Oseen's tensor. Greek indices indicate spatial components, and summation over repeated indices is implicit. In Fourier space, we have
\begin{equation} \label{eq Green-Fourier}
\tilde v_\alpha (\bm{q}) = \tilde G_{\alpha\beta}(\bm{q}) \tilde f_\beta(\bm{q}),
\end{equation}
where we have introduced the Fourier decomposition as
\begin{equation} \label{eq Fourier-flow}
\bm{v}(\bm{r}) = \int \frac{\dd^2 \bm{q}}{(2\pi)^2} \,\tilde{\bm{v}}(\bm{q}) \,e^{i \bm{q}\cdot\bm{r}}.
\end{equation}

To obtain the Green function, we must obtain the viscous stress exerted by the water and oil layers on the active fluid film,
\begin{subequations} \label{eq viscous-force}
\begin{align}
\bm{f}_{\text{water}} (\bm{r}) &= -\eta_{\text{water}} \left.\frac{\partial \bm{u}_\parallel(\bm{r},z)}{\partial z}\right|_{z=0^-},\\
\bm{f}_{\text{oil}} (\bm{r}) &= \eta_{\text{oil}} \left.\frac{\partial \bm{u}_\parallel(\bm{r},z)}{\partial z}\right|_{z=0^+},
\end{align}
\end{subequations}
respectively. Here, $\bm{u}(\bm{r},z)$ is the three-dimensional flow field of the passive fluids, and the subscript $\parallel$ indicates the components along the active film's plane ($z=0$, see \cref{sketch}). The viscous flows in the passive layers obey the Stokes equation
\begin{subequations}
\begin{align}
\eta_{\text{water}} \nabla^2 \bm{u} - \bm{\nabla} p = \bm{0};&\qquad -H_{\text{water}} < z < 0,\\
\eta_{\text{oil}} \nabla^2 \bm{u} - \bm{\nabla} p = \bm{0};&\qquad 0 < z < H_{\text{oil}},
\end{align}
\end{subequations}
with $p$ the three-dimensional pressure. These flows are driven by the hydrodynamic coupling with the active film: $\bm{u}_\parallel(\bm{r},0^+) = \bm{u}_\parallel(\bm{r},0^-) = \bm{v}(\bm{r})$. As shown in Refs. \cite{Stone1995,Stone1995a}, if the film's flow is incompressible, $\bm{\nabla}\cdot \bm{v} = 0$, then the pressure in the fluid layers is uniform, $\bm{\nabla}p=\bm{0}$, and the out-of-plane component of the velocity vanishes everywhere, $u_z = 0$. Therefore, the layers' flow field is planar and harmonic; it obeys Laplace's equation:
\begin{subequations} \label{eq Laplace}
\begin{align}
\eta_{\text{water}} \nabla^2 \bm{u}_\parallel = \bm{0};&\qquad -H_{\text{water}} < z < 0,\\
\eta_{\text{oil}} \nabla^2 \bm{u}_\parallel = \bm{0};&\qquad 0 < z < H_{\text{oil}},
\end{align}
\end{subequations}

To obtain the Green function in Fourier space, as in \cref{eq Green-Fourier}, we solve \cref{eq Laplace} in terms of the planar Fourier modes of the flow field, $\tilde{\bm{u}}(\bm{q},z)$, which obey
\begin{subequations} \label{eq Laplace-Fourier}
\begin{align}
\eta_{\text{water}} (\partial_z^2 - q^2) \tilde{\bm{u}} = \bm{0};&\qquad -H_{\text{water}} < z < 0,\\
\eta_{\text{oil}} (\partial_z^2 - q^2) \tilde{\bm{u}} = \bm{0};&\qquad 0 < z < H_{\text{oil}}.
\end{align}
\end{subequations}
The water layer is in contact with a solid substrate at $z=-H_{\text{water}}$, where we assume a no-slip boundary condition: $\bm{u}_\parallel(\bm{r},-H_{\text{water}}) = \bm{0}$. Respectively, the oil layer is in contact with air at $z=H_{\text{oil}}$, where we assume a no-shear-stress boundary condition: $\partial_z \bm{u}_\parallel (\bm{r},z)|_{z=H_{\text{oil}}} = 0$. With these boundary conditions and $\bm{u}_\parallel(\bm{r},0^+) = \bm{u}_\parallel(\bm{r},0^-) = \bm{v}(\bm{r})$, the solutions to \cref{eq Laplace-Fourier} are
\begin{subequations} \label{eq layers-flow}
\begin{align}
&\begin{multlined}
\tilde{\bm{u}}_\parallel(\bm{q},z) = \left[\cosh(qz) + \coth(qH_{\text{water}}) \sinh(qz)\right] \tilde{\bm{v}}(\bm{q});\\
-H_{\text{water}} < z < 0,
\end{multlined}
\\
&\begin{multlined}
\tilde{\bm{u}}_\parallel(\bm{q},z) = \left[\cosh(qz) - \tanh(qH_{\text{oil}}) \sinh(qz)\right] \tilde{\bm{v}}(\bm{q});\\
0 < z < H_{\text{oil}}.
\end{multlined}
\end{align}
\end{subequations}
These solutions show that the flow penetrates into the oil and water layers to depths given by the inverse of the in-plane wave number, $1/q$, unless limited by the layers' thicknesses, $H_{\text{water}}$ and $H_{\text{oil}}$, as illustrated in \cref{sketch}. Introducing \cref{eq layers-flow} into \cref{eq viscous-force}, we obtain
\begin{subequations}
\begin{align}
\tilde{\bm{f}}_{\text{water}} (\bm{q}) &= -\eta_{\text{water}} \,q\, \coth(q H_{\text{water}}) \,\tilde{\bm{v}}(\bm{q}),\\
\tilde{\bm{f}}_{\text{oil}} (\bm{q}) &= -\eta_{\text{oil}} \,q\, \tanh(q H_{\text{oil}}) \,\tilde{\bm{v}}(\bm{q}).
\end{align}
\end{subequations}

Finally, introducing these results into the Fourier transform of \cref{eq nematic-Stokes}, and using the incompressibility condition $\bm{q}\cdot\tilde{\bm{v}} = 0$, we obtain the Green function
\begin{equation} \label{eq Green}
\tilde G_{\alpha\beta}(\bm{q}) = \frac{\delta_{\alpha\beta} - q_\alpha q_\beta /q^2}{\eta_{\text{n}} q^2 + \eta_{\text{oil}} q \tanh(qH_{\text{oil}}) + \eta_{\text{water}} q \coth(qH_{\text{water}})}.
\end{equation}
This function describes the hydrodynamic interactions in the active nematic film, accounting both for the direct interactions due to its incompressibility and viscosity, and also for the indirect interactions mediated by the the oil and water layers. The ratios between the viscosity of these external fluids and the two-dimensional viscosity of the nematic film define two viscous length scales:
\begin{equation}
\ell_{\text{oil}} = \eta_{\text{n}}/\eta_{\text{oil}},\qquad \ell_{\text{water}} = \eta_{\text{n}}/\eta_{\text{water}}.
\end{equation}
At scales larger than this viscous length, dissipation in the external fluid layer (either oil or water) dominates over dissipation within the nematic film.

\subsection{Kinetic energy spectrum}
\label[SI section]{spectrum derivation}

In this section, we derive an analytical expression for the kinetic energy spectrum of the turbulent flows in our system. To this end, we establish a relationship between the flow field of our coupled three-fluid system (\cref{sketch}) and the vorticity field of an isolated active nematic film. By means of this relationship, we combine the hydrodynamic Green function obtained in \cref{Green's function} with Giomi's mean-field theory of active nematic turbulence \cite{Giomi2015} to predict the flow power spectrum in our experimental system.

The kinetic energy per unit mass density, $E$, of the active nematic flows is given by
\begin{equation}
E = \frac{1}{2} \int \bm{v}^2 \,\dd^2\bm{r}.
\end{equation}
Using the Fourier modes $\tilde{\bm{v}}(\bm{q})$ of the flow field, as introduced in \cref{eq Fourier-flow}, the angle-averaged spectrum $E(q)$, with $q=|\bm{q}|$ the wave number, is defined by
\begin{equation}
\langle E\rangle = \frac{1}{2} \int \frac{\dd^2\bm{q}}{(2\pi)^2} \langle |\tilde{\bm{v}}(\bm{q})|^2 \rangle = \mathcal{A} \int_0^\infty E(q) \dd q,
\end{equation}
where $\mathcal{A}$ is the area of the system, and $\langle \cdot \rangle$ averages over realizations. In states where correlations of the flow field are isotropic, $E(q)$ is given by
\begin{equation} \label{eq kinetic-energy-spectrum-definition}
E(q) = \frac{1}{4\pi \mathcal{A}} q\, \langle \left|\tilde{\bm{v}} (\bm{q})\right|^2 \rangle.
\end{equation}

To obtain the velocity power spectrum, we use \cref{eq Green-Fourier,eq Green}, which give
\begin{multline} \label{eq velocity-spectrum}
\langle |\tilde{\bm{v}} (\bm{q}) |^2\rangle = \left(\delta_{\alpha\beta} - \frac{q_\alpha q_\beta}{q^2}\right) \left(\delta_{\alpha\gamma} - \frac{q_\alpha q_\gamma}{q^2}\right) \frac{\langle \tilde f_\beta (\bm{q}) \tilde f_\gamma^* (\bm{q})\rangle}{ \Lambda^2 (q) } \\
= \frac{1}{ \Lambda^2 (q) } \left\langle |\tilde{\bm{f}}(\bm{q})|^2 - \frac{q_\alpha q_\beta}{q^2} \tilde f_\alpha(\bm{q}) \tilde f_\beta^*(\bm{q}) \right\rangle \\
= \frac{1}{ q^2 \Lambda^2 (q) } \left\langle q_y^2 |\tilde f_x |^2 + q_x^2 |\tilde f_y |^2 - q_x q_y (\tilde f_x \tilde f_y^* + \tilde f_y \tilde f_x^*) \right\rangle.
\end{multline}
Here, we have introduced the notation
\begin{equation} \label{eq hydrodynamic-kernel}
\Lambda(q) \equiv \eta_{\text{n}} q^2 + \eta_{\text{oil}} q \tanh(qH_{\text{oil}}) + \eta_{\text{water}} q \coth(qH_{\text{water}}).
\end{equation}

To eliminate the source force density $\bm{f}$, and thereby obtain a closed-form expression for the velocity power spectrum, we leverage the force-balance condition for the active nematic film alone, without external fluids:
\begin{equation} \label{eq Stokes-alone}
\eta_{\text{n}} \nabla^2 \bm{v}_{\text{i}} - \bm{\nabla}P_{\text{i}} + \bm{f} = \bm{0}.
\end{equation}
Here, the subscript i indicates that the active nematic film is isolated. As in \cref{eq nematic-Stokes}, $\bm{f}$ accounts for source force density due to elastic, flow-alignment, and active stresses, albeit now in the isolated film. The source force of the isolated film (in \cref{eq Stokes-alone}) coincides with that of the full problem with external fluid layers (in \cref{eq nematic-Stokes}) only if we ignore flow-alignment effects. In this limit, $\bm{f}$ does not depend explicitly on the flow field, and hence it is insensitive to the presence of external fluid layers. To exploit this fact and therefore be able to
derive a closed form for the velocity power spectrum, we first ignore {the flow alignment coupling.

Following Ref. \cite{Alert2020b}, we take the curl of \cref{eq Stokes-alone} to obtain a Poisson equation for the vorticity field $\omega = \hat{\bm{z}}\cdot (\bm{\nabla}\times \bm{v})$:
\begin{equation}
\nabla^2 \omega_{\text{i}} = s(\bm{r},t);\qquad s(\bm{r},t) = -\frac{1}{\eta_{\text{n}}}\hat{\bm{z}}\cdot (\bm{\nabla}\times \bm{f}),
\end{equation}
where $s$ is the vorticity source due to nematic forces. In Fourier space, this equation is written as
\begin{equation}
-q^2 \tilde \omega_{\text{i}}(\bm{q}) = \tilde s(\bm{q}) = \frac{i}{\eta_{\text{n}}} ( q_y \tilde f_x - q_x \tilde f_y).
\end{equation}
Hence, the vorticity spectrum of the isolated active film is given by
\begin{equation}
\langle |\tilde \omega_{\text{i}}(\bm{q})|^2 \rangle = \frac{1}{\eta_{\text{n}}^2 q^4} \left\langle q_y^2 |\tilde f_x |^2 + q_x^2 |\tilde f_y |^2 - q_x q_y (\tilde f_x \tilde f_y^* + \tilde f_y \tilde f_x^*) \right\rangle.
\end{equation}
Comparing to \cref{eq velocity-spectrum}, we obtain
\begin{equation} \label{eq velocity-vorticity-spectra}
\langle |\tilde{\bm{v}} (\bm{q}) |^2\rangle = \frac{\eta_{\text{n}}^2 q^2}{\Lambda^2(q)} \langle |\tilde \omega_{\text{i}}(\bm{q})|^2 \rangle,
\end{equation}
with $\Lambda(q)$ given by \cref{eq hydrodynamic-kernel}. As explained above, this result is exact in the limit of vanishing flow alignment coupling. In the presence of flow alignment, \cref{eq velocity-vorticity-spectra} defines a closed approximation whereby the nematic forces, both passive and active, are included exactly, while the flow alignment forces are approximated by the hydrodynamics of the isolated problem.

\Cref{eq velocity-vorticity-spectra} relates the velocity spectrum of the active nematic film coupled to the external fluid layers with the vorticity spectrum of an isolated active nematic film. For the latter, we can now make use of the mean-field theory introduced by Giomi \cite{Giomi2015}}, which is based on decomposing the vorticity field into a superposition of $N$ uncorrelated vortices. Based on simulation results, the theory assumes that each vortex has a vorticity $\omega_{\text{v}}$ independent of its size, and that vortex areas follow an exponential distribution with mean $a_* = \pi R_*^2$, where $R_*$ is the mean vortex radius. With these assumptions, Giomi's mean-field theory predicts \cite{Giomi2015}
\begin{equation} \label{eq vorticity-spectrum}
\langle |\tilde \omega_{\text{i}}(\bm{q})|^2\rangle = \frac{N \omega_{\text{v}}^2 R_*^4 }{8 \pi^2} e^{-q^2 R_*^2/2} \left[ I_0\left(\frac{q^2 R_*^2}{2}\right) - I_1\left(\frac{q^2 R_*^2}{2}\right)\right],
\end{equation}
where $I_0$ and $I_1$ are modified Bessel functions of the first kind. Introducing this result into \cref{eq velocity-vorticity-spectra}, and into \cref{eq kinetic-energy-spectrum-definition}, we obtain the kinetic energy spectrum
\begin{equation} \label{eq energy-spectrum}
E(q) = \frac{B q R_*^4\, e^{-q^2 R_*^2/2} \left[ I_0\left(q^2 R_*^2/2\right) - I_1\left(q^2 R_*^2/2\right)\right]}{\left[q + \eta_{\text{oil}}/\eta_{\text{n}} \tanh(qH_{\text{oil}}) + \eta_{\text{water}}/\eta_{\text{n}} \coth(qH_{\text{water}})\right]^2},
\end{equation}
where $B = N \omega^2_{\text{v}}/(32\pi^3 \mathcal{A})$ is a prefactor related to the total enstrophy, and independent of both the wave number $q$ and the mean vortex radius $R_*$.

\subsection{Scaling laws} \label[SI section]{scaling-laws}

\begin{figure}[tb]
\begin{center}
    \includegraphics[width=\columnwidth]{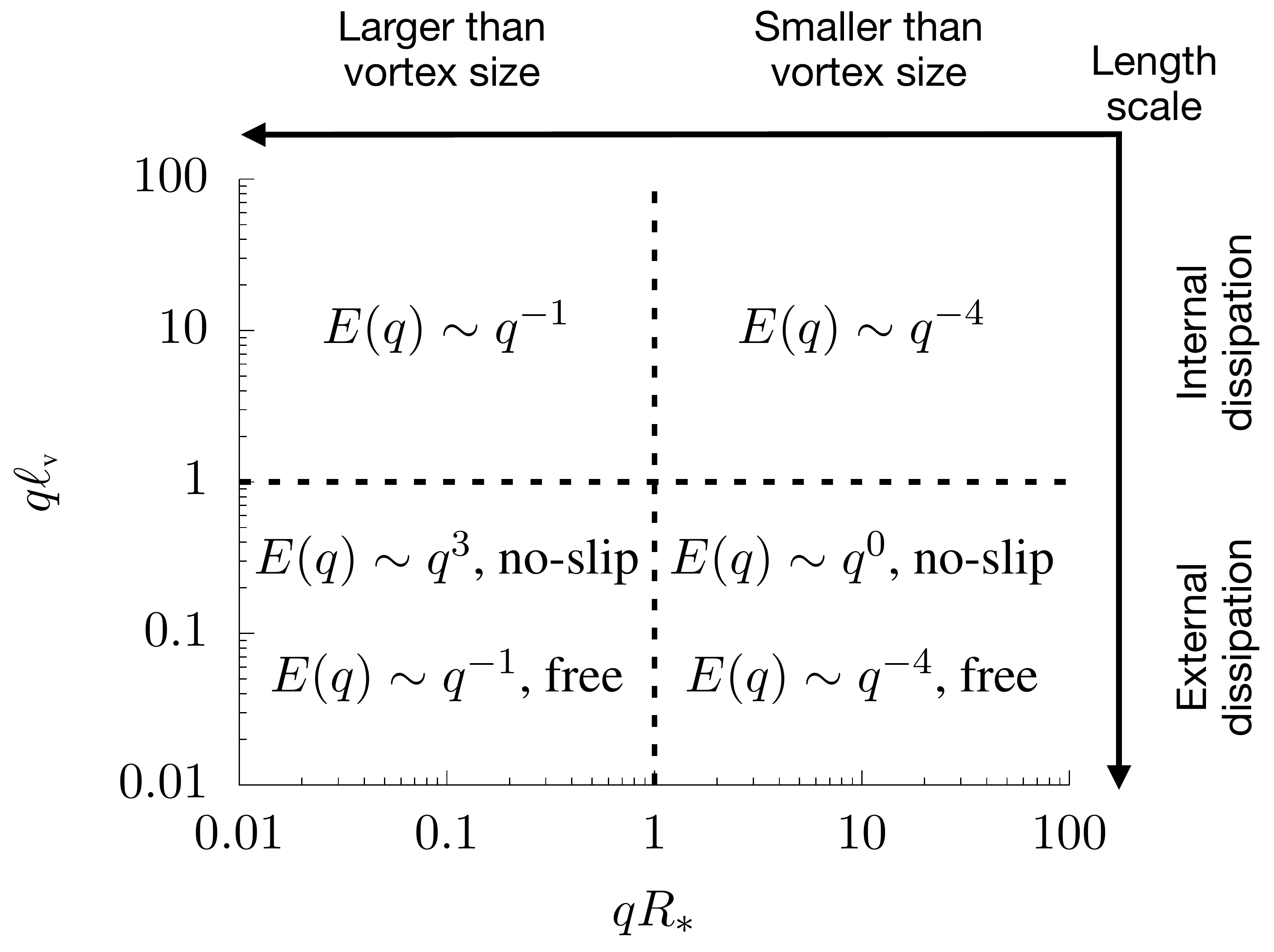}
\end{center}
\bfcaption{Scaling regimes of turbulent flows in an active nematic film in contact with an external fluid}{ The different regimes are predicted at length scales ($\sim 1/q$) either larger or smaller than the mean vortex radius $R_*$, the viscous length $\ell_{\text{v}} = \eta_{\text{n}}/\eta_{\text{ext}}$, and the thickness $H$ of the external fluid layer. This figure summarizes the scalings in the thin-layer limit $q H\ll 1$; see the main text and \cref{Fig2} for the predictions in the thick-layer limit $q H\gg 1$.}
\label{Fig thin}
\end{figure}

\begin{figure*}[ptbh]
\begin{center}
    \includegraphics[width=0.65\textwidth]{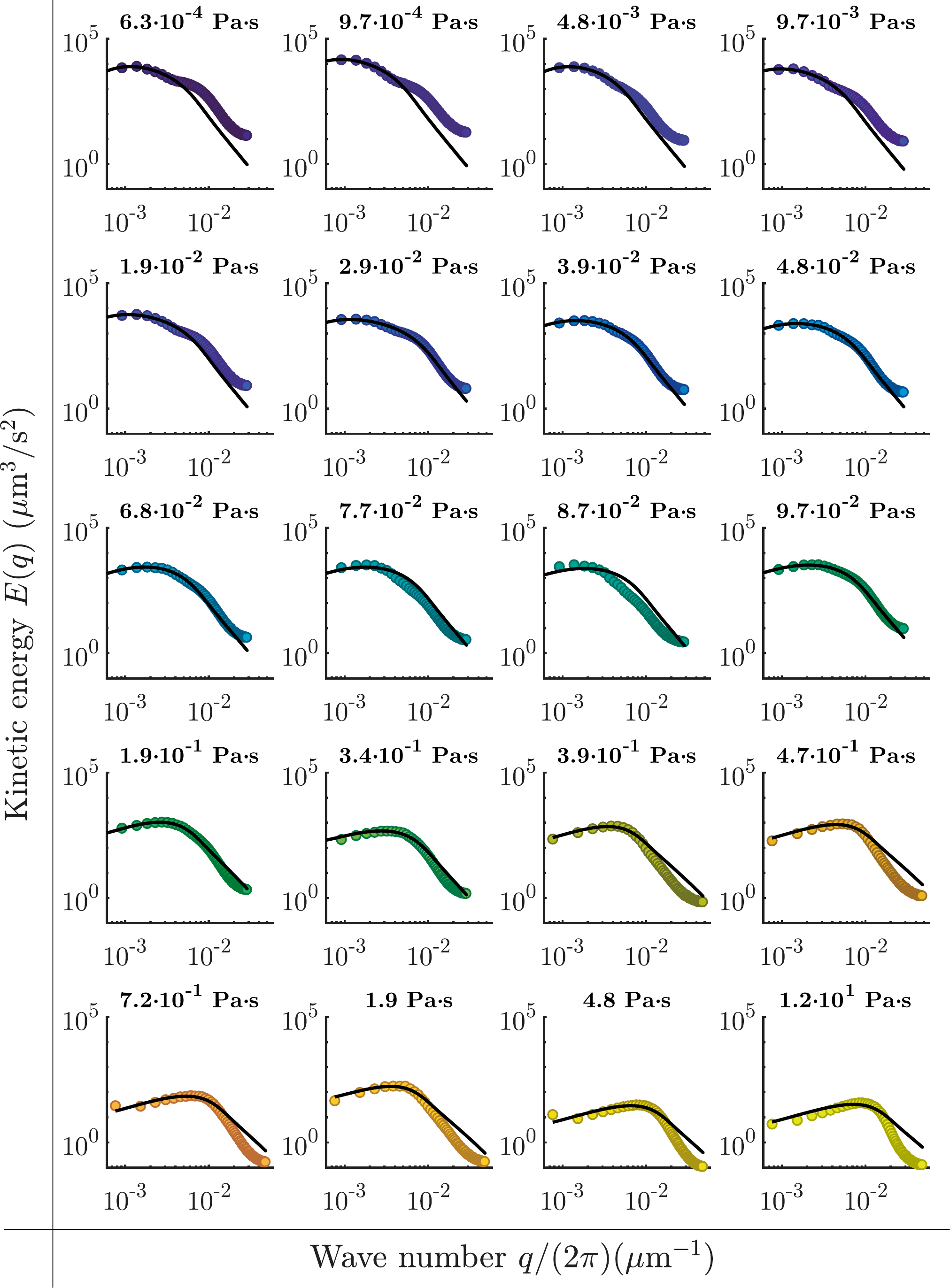}
\end{center}
\bfcaption{Fits of the spectra for all oil viscosities}{ The theory (\cref{eq energy-spectrum-main}) fits the data well for an intermediate range of oil viscosities, as exemplified in \cref{Fig fit}, with departures for both very low and very high oil viscosities.}
\label{Fig spectra-all-viscosities}
\end{figure*}

\begin{figure*}[tbh]
\begin{center}
    \includegraphics[width=0.35\textwidth]{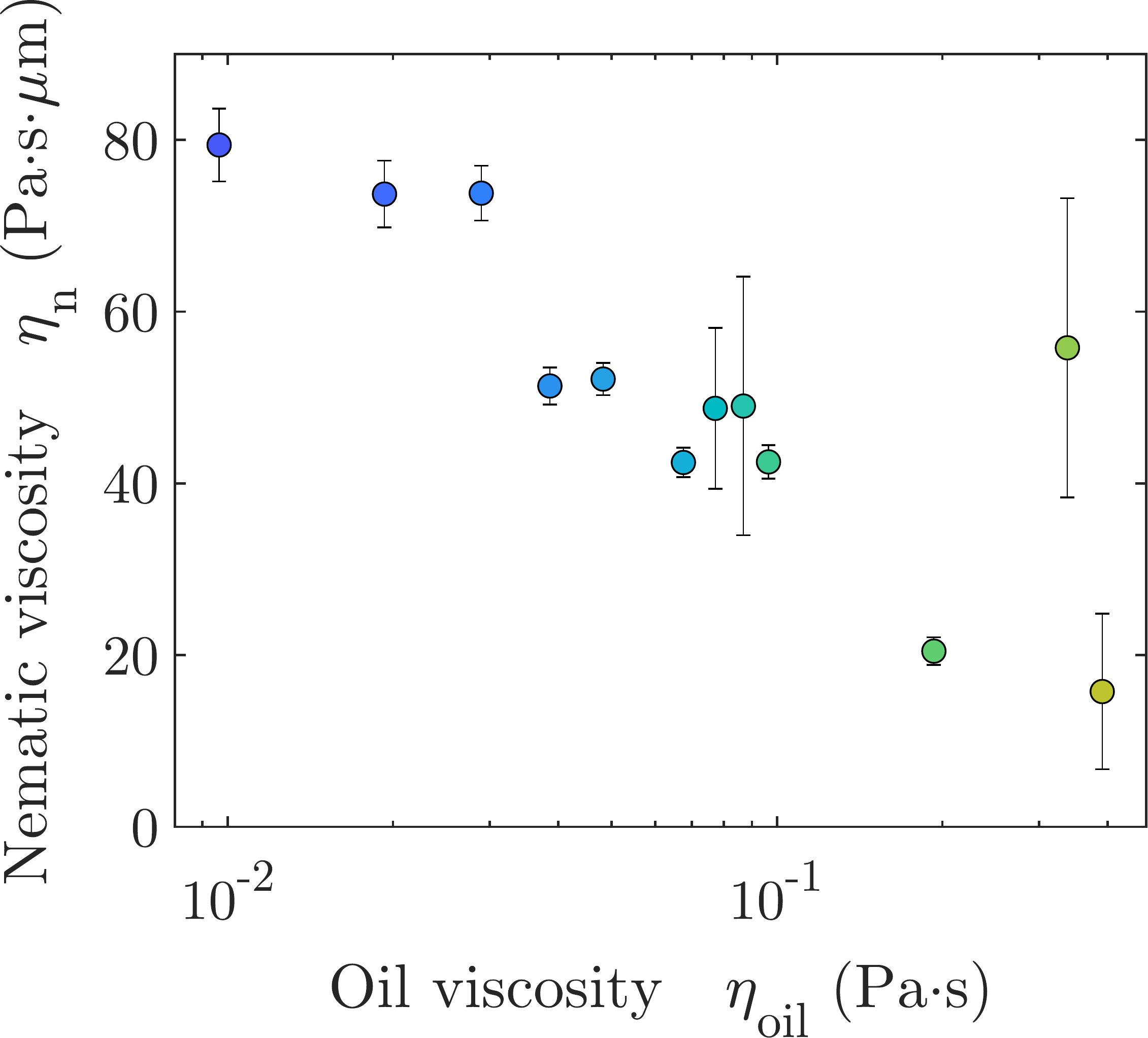}
\end{center}
\bfcaption{Active nematic viscosity}{ Effective viscosity of the active nematic layer, $\eta_{\text{n}}$, in contact with oils of different viscosities. We obtained $\eta_{\text{n}}$ from the fits of \cref{eq energy-spectrum-main} to the experimental data, as in \cref{Fig fit}.}
\label{Fig nematic-viscosity}
\end{figure*}

In this section, we extend the discussion of the predicted scaling regimes given in the main text. As in the main text, we consider a situation with just one external fluid layer and classify the scaling regimes in terms of three characteristic lengths: the mean vortex radius $R_*$, the viscous length $\ell_{\text{v}} = \eta_{\text{n}}/\eta_{\text{ext}}$, and the external layer thickness $H$. In the main text, we discuss the scaling regimes at scales much smaller than the layer thickness, $qH \gg 1$, summarizing our results in \cref{Fig2}. Here, we give the results in the opposite limit, looking at scales much larger than the layer thickness, $q H\ll 1$. In this thin-layer limit, the scaling laws depend on the boundary condition of the external fluid at the non-active interface. We considered both a no-slip boundary condition, as for the water-substrate interface in our experiments, and a free surface, as for the oil-air interface in our experiments (\cref{sketch}). We summarize the results for all these situations in \cref{Fig thin}, which shows two additional scaling regimes, $E(q)\sim q^3$ and $E(q)\sim q^0$, for no-slip boundary conditions.

\subsection{Vortex size selection} \label[SI section]{selection}

In this section, we discuss the physical origin of the mean vortex size $R_*$, a key parameter in the description of active turbulent flows. Active nematic turbulence results from the well-known spontaneous-flow instability in active nematic fluids \cite{Simha2002,Voituriez2005,Marchetti2013}. Here, we ask whether vortex size is selected at the early stages of the instability, i.e. by its linear dynamics, or later on by the nonlinear dynamics.

In isolated active nematic films, the spontaneous-flow instability is a long-wavelength instability, whose growth rate is maximal at the longest wavelengths. As a result, the linear regime of the instability does not select any intrinsic wavelength but rather produces flow patterns with wavelengths determined by the system size. Indeed, system-size-dependent, stationary vortex patterns were observed in simulations at small activities, just past the instability threshold \cite{Alert2020b}. At high activity, however, these stationary vortices were unstable and evolved into turbulent flows. Simulations without flow-alignment revealed a sequence of instabilities whereby vortices break into smaller vortices, down to a characteristic vortex size proportional to the critical wavelength of the instability, $\lambda_{\text{c}}$, which is itself proportional to the active length $\ell_{\text{a}}$ \cite{Alert2020b}. In the presence of flow alignment, a similar transient cascade may exist. Moreover, flow alignment introduces a new nonlinearity in the problem enabling an additional potential mechanism of energy transfer in the steady state. It is thus reasonable to expect that, also in the presence of flow alignment, the vortex size for an isolated nematic is selected by a nonlinear mechanism.

This scenario is modified when the active nematic is not isolated but coupled either to a frictional substrate \cite{Thampi2014,Duclos2018} or to external fluids \cite{Sarkar2015,Gao2015,Gao2017,Martinez-Prat2019}, as in our system. The external fluids are isotropic and in the Stokes regime, and therefore they cannot transfer energy across scales at the steady state. However, the coupling to an external fluid modifies the spontaneous-flow instability of the active nematic at the linear level, making the growth rate achieve a maximum at finite wavelengths \cite{Gao2015,Gao2017}. Consequently, an intrinsic wavelength $\lambda_{\text{m}}$ is selected from the very onset of spontaneous flows, which dominates the early stages of turbulence development \cite{Martinez-Prat2019}. To what extent nonlinear effects may modify this characteristic scale at later stages of turbulence is an open question that we can now analyze in light of our data.

To this end, we derive the wavelength $\lambda_{\text{m}}$ that is selected in the linear regime by the coupling between the active nematic and the external fluids in our experiments. We then compare these predictions to our experimental measurements of the vortex size in fully-developed turbulence at different oil viscosities.

To obtain the selected wavelength, we analyze the dynamics around the uniformly-oriented quiescent state of the active nematic film. Ignoring flow alignment and topological defects, the dynamics of the angle $\theta$ of the nematic director field $\hat{\bm{n}} = (\cos\theta,\sin\theta)$ can be written as \cite{Alert2020b}
\begin{equation} \label{eq angle-dynamics}
\partial_t \theta + \bm{v}\cdot \bm{\nabla}\theta + \frac{\omega}{2} = \frac{K}{\gamma} \nabla^2\theta.
\end{equation}
The right-hand side of this equation accounts for the relaxation of distortions of the director field, which generate elastic nematic stress. We describe this stress in the approximation of one Frank constant $K$ which, together with the rotational viscosity $\gamma$, controls the director relaxation rate \cite{deGennes-Prost}. Respectively, the second and third terms on the left-hand side of \cref{eq angle-dynamics} account for the advection and corotation of the nematic director by the flow field $\bm{v}$, with vorticity $\omega = \hat{\bm{z}}\cdot (\bm{\nabla}\times \bm{v})$.

In turn, the flow field is driven by nematic forces $\bm{f}$ as specified by the force balance \cref{eq nematic-Stokes}. Ignoring flow alignment as in \cref{eq angle-dynamics}, the nematic forces are given by
\begin{equation} \label{eq nematic-force}
f_\alpha = \partial_\beta (\sigma_{\alpha\beta}^{\text{ant}} + \sigma_{\alpha\beta}^{\text{act}}).
\end{equation}
The first contribution accounts for elastic nematic stresses described by the antisymmetric part of the the stress tensor \cite{deGennes-Prost},
\begin{equation}
\sigma_{\alpha\beta}^{\text{ant}} = \frac{1}{2} (n_\alpha h_\beta - n_\beta h_\alpha).
\end{equation}
Here, $\bm{h} = -\delta F_{\text{n}}/\delta \hat{\bm{n}} = K\nabla^2 \hat{\bm{n}}$ is the molecular field computed from the Frank free energy for nematic elasticity which, in the one-constant approximation, is given by \cite{deGennes-Prost}
\begin{equation}
F_{\text{n}} = \frac{K}{2}\int (\partial_\alpha n_\beta)(\partial_\alpha n_\beta)\,\dd^2\bm{r} = \frac{K}{2} \int |\bm{\nabla} \theta|^2\, \dd^2\bm{r}.
\end{equation}
Respectively, the second contribution in \cref{eq nematic-force} corresponds to the active nematic stress \cite{Kruse2005,Marchetti2013,Prost2015,Julicher2018}
\begin{equation}
\sigma_{\alpha\beta}^{\text{act}} = -\zeta q_{\alpha\beta},
\end{equation}
where $\zeta$ is the active stress coefficient, and $q_{\alpha\beta} = n_\alpha n_\beta - 1/2\,\delta_{\alpha\beta}$ is the nematic orientation tensor in two dimensions, with cartesian components

\begin{subequations} \label[equation]{eq q}
\begin{align}
q_{xx} &= -q_{yy} = \frac{1}{2}\cos(2\theta),\\
q_{xy} &= q_{yx} = \frac{1}{2}\sin(2\theta).
\end{align}
\end{subequations}
To perform a linear stability analysis, we introduce perturbations around the reference state as $\theta = 0 + \delta\theta$ and $\bm{v} = \bm{0} + \delta\bm{v}$, and we decompose them into Fourier-Laplace modes as

\begin{subequations}
\begin{align}
\delta \theta(\bm{r},t) &= \int \frac{\dd \Omega}{2\pi} \int \frac{\dd^2\bm{q}}{(2\pi)^2} \,\delta\tilde\theta (\bm{q},\Omega)\, e^{\Omega t + i \bm{q}\cdot\bm{r}},\\
\delta \bm{v}(\bm{r},t) &= \int \frac{\dd \Omega}{2\pi} \int \frac{\dd^2\bm{q}}{(2\pi)^2} \,\delta\tilde{\bm{v}} (\bm{q},\Omega)\, e^{\Omega t + i \bm{q}\cdot\bm{r}},
\end{align}
\end{subequations}
with wave vector $\bm{q}$ and frequency $\Omega$. In terms of these modes, and to first order in perturbations, \cref{eq angle-dynamics} is written as

\begin{equation} \label{eq angle-Fourier}
\Omega \,\delta\tilde\theta + \frac{1}{2} ( i q_y \delta\tilde v_x - i q_x \delta\tilde v_x) = -\frac{K}{\gamma} q^2 \delta\tilde\theta.
\end{equation}
In turn, the Fourier modes of the flow field are related to those of the nematic force, $\tilde{\bm{f}}$, via \cref{eq Green-Fourier,eq Green}. To first order in perturbations, \crefrange{eq nematic-force}{eq q} yield

\begin{subequations}
\begin{align}
\delta \tilde f_x & = i q_y \left(-\zeta - \frac{K}{2} q^2\right) \delta\tilde\theta,\\
\delta \tilde f_y & = i q_x \left(-\zeta + \frac{K}{2} q^2\right) \delta\tilde\theta.
\end{align}
\end{subequations}
Introducing these results into \cref{eq Green-Fourier}, using \cref{eq Green}, and then introducing the resulting $\delta\tilde{\bm{v}}$ into \cref{eq angle-Fourier}, we obtain the growth rate of perturbations:
\begin{equation} \label{eq growth-rate}
\Omega(\bm{q}) = - \frac{K}{\gamma} q^2 + \frac{\zeta q^2 \cos(2\phi)/2 - K q^4 /4}{\Lambda(q)},
\end{equation}
where $\phi$ is the angle formed by the wave vector $\bm{q}$ and the director $\hat{\bm{n}}$, such that $\bm{q}\cdot \hat{\bm{n}} = q \cos\phi$, and $\Lambda(q)$ is the hydrodynamic kernel given by \cref{eq hydrodynamic-kernel}.

To get insight into wavelength selection, we consider a situation with only one external fluid layer with thickness $H\rightarrow \infty$ and viscous length $\ell_{\text{v}} = \eta_{\text{n}}/\eta_{\text{ext}}$. In this situation, \cref{eq growth-rate} takes the simpler form
\begin{equation} \label{eq simple-growth-rate}
\Omega(\bm{q}) = - \frac{K}{\gamma} q^2 + \frac{1}{\eta_{\text{ext}}} \frac{\zeta q^2 \cos(2\phi)/2 - K q^4 /4}{\ell_{\text{v}} q^2 + q}.
\end{equation}
Rescaling length and time by the active length and time, respectively,
\begin{equation} \label{eq active-length-time}
\ell_{\text{a}} = \sqrt{\frac{K}{|\zeta|} \frac{\eta_{\text{n}}}{\gamma}},\qquad \tau_{\text{a}} = \frac{\gamma}{K}\ell_{\text{a}}^2 = \frac{\eta_{\text{n}}}{|\zeta|},
\end{equation}
the growth rate \cref{eq simple-growth-rate} can be expressed in dimensionless form as
\begin{equation} \label{eq dimensionless-simple-growth-rate}
\bar\Omega(q,\phi) = -\bar q^2 + \frac{ \sgn(\zeta) \bar q^2 \cos(2\phi)/2 - r \bar q^4/4}{ \bar q^2 + \bar q / \bar\ell_{\text{v}}}.
\end{equation}
Here, $\bar\Omega = \Omega \tau_{\text{a}}$, and $\bar q = q \ell_{\text{a}}$ are dimensionless variables. We have also introduced three dimensionless parameters: the viscosity ratio $r\equiv \gamma/\eta_{\text{n}}$, the dimensionless viscous length $\bar\ell_{\text{v}} = \ell_{\text{v}} / \ell_{\text{a}}$, and the sign of the active stress $\sgn(\zeta) =\pm 1$ for extensile and contractile stresses, respectively.

\begin{figure*}[tbh!]
\begin{center}
    \includegraphics[width=\textwidth]{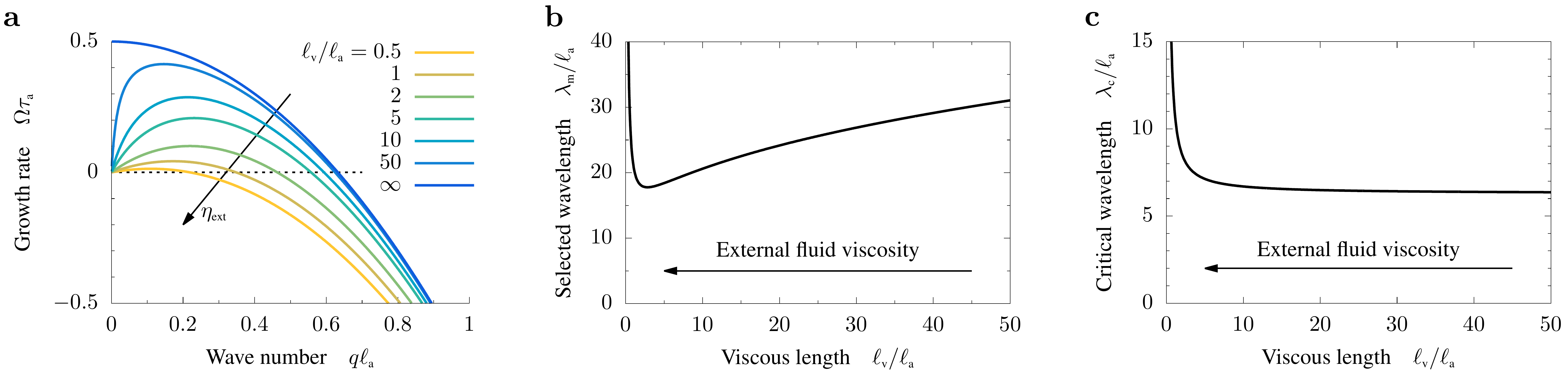}
  {\phantomsubcaption\label{growth-rate}}
  {\phantomsubcaption\label{selected-wavelength}}
  {\phantomsubcaption\label{critical-wavelength}}
  {\phantomsubcaption\label{vortex-radius}}
\end{center}
\bfcaption{Vortex size selection}{ \subref*{growth-rate}, Growth rate (\cref{eq dimensionless-simple-growth-rate}) along the most unstable direction for the spontaneous-flow instability in an active nematic film coupled to an external fluid layer. The viscosity ratio is set to $r=\gamma/\eta_{\text{n}} = 1$, and flow alignment is ignored. The different curves correspond to different values of the external fluid viscosity $\eta_{\text{ext}}$, expressed in terms of the viscous length $\ell_{\text{v}} = \eta_{\text{n}}/\eta_{\text{ext}}$. The blue line corresponds to an isolated active nematic film, without external fluid. Lengths and time are rescaled by the active length and time, $\ell_{\text{a}}$ and $\tau_{\text{a}}$ defined in \cref{eq active-length-time}. \subref*{selected-wavelength}, Wavelength at which the growth rate is maximum, as obtained from \cref{eq maximum}, as a function of the viscous length. This wavelength is selected by the linear dynamics upon the spontaneous-flow instability. \subref*{critical-wavelength}, Critical wavelength of the spontaneous-flow instability (\cref{eq critical}) as a function of the viscous length.}
\label{vortex-size-selection}
\end{figure*}

The direction of most unstable perturbations is along the director ($\phi^* = 0$) for extensile stresses ($\zeta>0$), and perpendicular to the director ($\phi^*=\pi/2$) for contractile stresses ($\zeta<0$). Along the most unstable direction, the growth rate \cref{eq dimensionless-simple-growth-rate} has the shape shown in \cref{growth-rate}. This figure shows that the coupling of the active nematic to the external fluid layer, represented by a finite $\bar\ell_{\text{v}} = \ell_{\text{v}}/\ell_{\text{a}}$, produces a maximum of the growth rate at a finite wavelength. This selected wavelength, $\lambda_{\text{m}} = 2\pi/q_{\text{m}}$, is obtained from the single real solution of the following cubic equation:
\begin{equation} \label{eq maximum}
\left(2 + \frac{r}{2}\right) \bar\ell_{\text{v}}^2 \bar q_{\text{m}}^3 + \left(4 + \frac{3r}{4}\right) \bar\ell_{\text{v}} \bar q_{\text{m}}^2 + 2 \bar q_{\text{m}} - \frac{1}{2}\bar\ell_{\text{v}} = 0.
\end{equation}
As shown in \cref{selected-wavelength}, and also apparent in \cref{growth-rate}, the selected wavelength $\lambda_{\text{m}}$ has a non-monotonic dependence on the viscous length $\ell_{\text{v}}$. In the common situation in which the viscous length is larger than the active length, the selected wavelength decreases with the viscosity of the external fluid. However, when the viscous length becomes smaller than the active length, the selected wavelength increases with the viscosity of the external fluid.

The coupling to the external fluid not only gives rise to the selected wavelength $\lambda_{\text{m}}$ but also modifies the critical wavelength of the instability, $\lambda_{\text{c}} = 2\pi/q_{\text{c}}$. This wavelength is determined by the condition $\Omega(q_{\text{c}}) = 0$ along the most unstable direction, which gives
\begin{equation} \label{eq critical}
q_{\text{c}} = \frac{1}{(2+r/2) \ell_{\text{v}}} \left[ -1 + \sqrt{1 + (2+r/2) \frac{\ell_{\text{v}}^2}{\ell_{\text{a}}^2}}\right],
\end{equation}
with $\ell_{\text{a}}$ given in \cref{eq active-length-time}. As shown in \cref{critical-wavelength}, the critical wavelength $\lambda_{\text{c}}$ increases monotonically with the viscosity $\eta_{\text{ext}}$ of the external fluid. At small $\eta_{\text{ext}}$, i.e. large $\ell_{\text{v}} = \eta_{\text{n}}/\eta_{\text{ext}}$, the critical wavelength saturates at its value for isolated nematic films:
\begin{equation}
\lim_{\ell_{\text{v}} \gg \ell_{\text{a}}} \lambda_{\text{c}} = 2\pi \sqrt{2 + r/2}\, \ell_{\text{a}}.
\end{equation}
In the opposite limit, when dissipation is dominated by the viscosity of the external fluid, the critical wavelength becomes proportional to the external fluid viscosity:
\begin{equation}
\lim_{\ell_{\text{v}} \ll \ell_{\text{a}}} \lambda_{\text{c}} = 4\pi \ell_{\text{a}} \frac{1}{\bar\ell_{\text{v}}} = 4\pi \ell_{\text{a}}^2 \frac{\eta_{\text{ext}}}{\eta_{\text{n}}}.
\end{equation}

In summary, the coupling to the external fluid endows the spontaneous flow instability with a linear wavelength-selection mechanism. This mechanism selects the characteristic scale of turbulent flows at early stages \cite{Martinez-Prat2019}. However, nonlinear effects may modify the selected scale at later stages, and thereby dictate the vortex size in the stationary, fully-developed turbulent regime. To assess this point in our experiments, we analyze how stationary vortex size varies with oil viscosity.

If stationary vortex size were dictated by the linear wavelength-selection mechanism, vortex size should vary with oil viscosity as in \cref{selected-wavelength}. In our experiments, the oil viscous length is $\ell_{\text{oil}} \gtrsim R_*$, and the vortex size should be $R_*\gtrsim \ell_{\text{a}}$. Therefore, we are in the regime $\ell_{\text{v}}>\ell_{\text{a}}$, and vortex size should moderately decrease with oil viscosity (\cref{selected-wavelength}). Instead, if vortex size were dictated by the nonlinear selection mechanism of Ref. \cite{Alert2020b}, vortex size should be proportional to the critical wavelength, and therefore vary with oil viscosity as in \cref{critical-wavelength}. Thus, again in the regime $\ell_{\text{v}}>\ell_{\text{a}}$, vortex size should remain roughly independent of oil viscosity (\cref{critical-wavelength}).

Our experimental measurements show that the mean vortex radius, as measured from the vortex area distribution, is rather independent of oil viscosity over a range spanning several orders of magnitude of oil viscosity (\cref{Fig vortex-radius-distribution}). At very high oil viscosities, we observe a moderate decrease of vortex radius (\cref{Fig vortex-radius-distribution}), whose interpretation remains an open question.

This analysis suggests that the linear selection mechanism alone does not explain the observed vortex size in the stationary regime, thus implying some form of energy transfer across scales, be it transient and/or steady. Elucidating the mechanism of vortex size selection in more detail thus remits to fundamental open questions in active turbulence that we defer to future work.

\onecolumngrid
\clearpage

\section*{Supplementary Figures}
\label{SFig}

\begin{figure*}[tbh]
\begin{center}
    \includegraphics[width=0.35\textwidth]{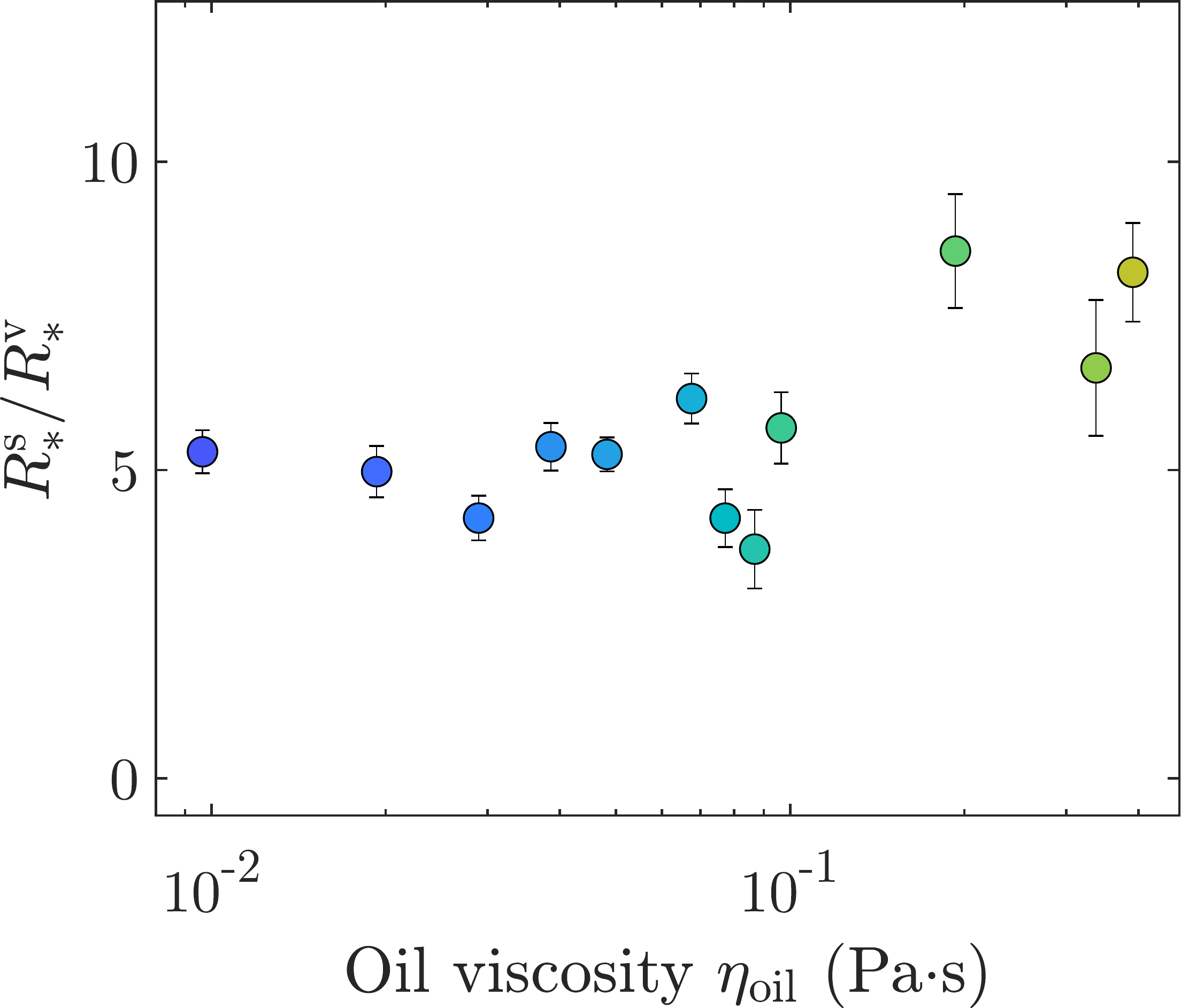}
\end{center}
\bfcaption{Comparing vortex sizes}{ Ratio of the mean vortex radii obtained from the spectral fits ($R^{\text{s}}_*$, \cref{Fig vortex-radius-spectra}) and from the vortex area distribution fits ($R^{\text{v}}_*$, \cref{Fig vortex-radius-distribution}). In the range of oil viscosities in which the theory fits the data well, both estimates of $R_*$ exhibit similar trends with oil viscosity. Hence, their ratio is rather independent of oil viscosity.}
\label{Fig vortex-radii}
\end{figure*}

\begin{figure*}[tbhh]
\begin{center}
    \includegraphics[width=0.45\textwidth]{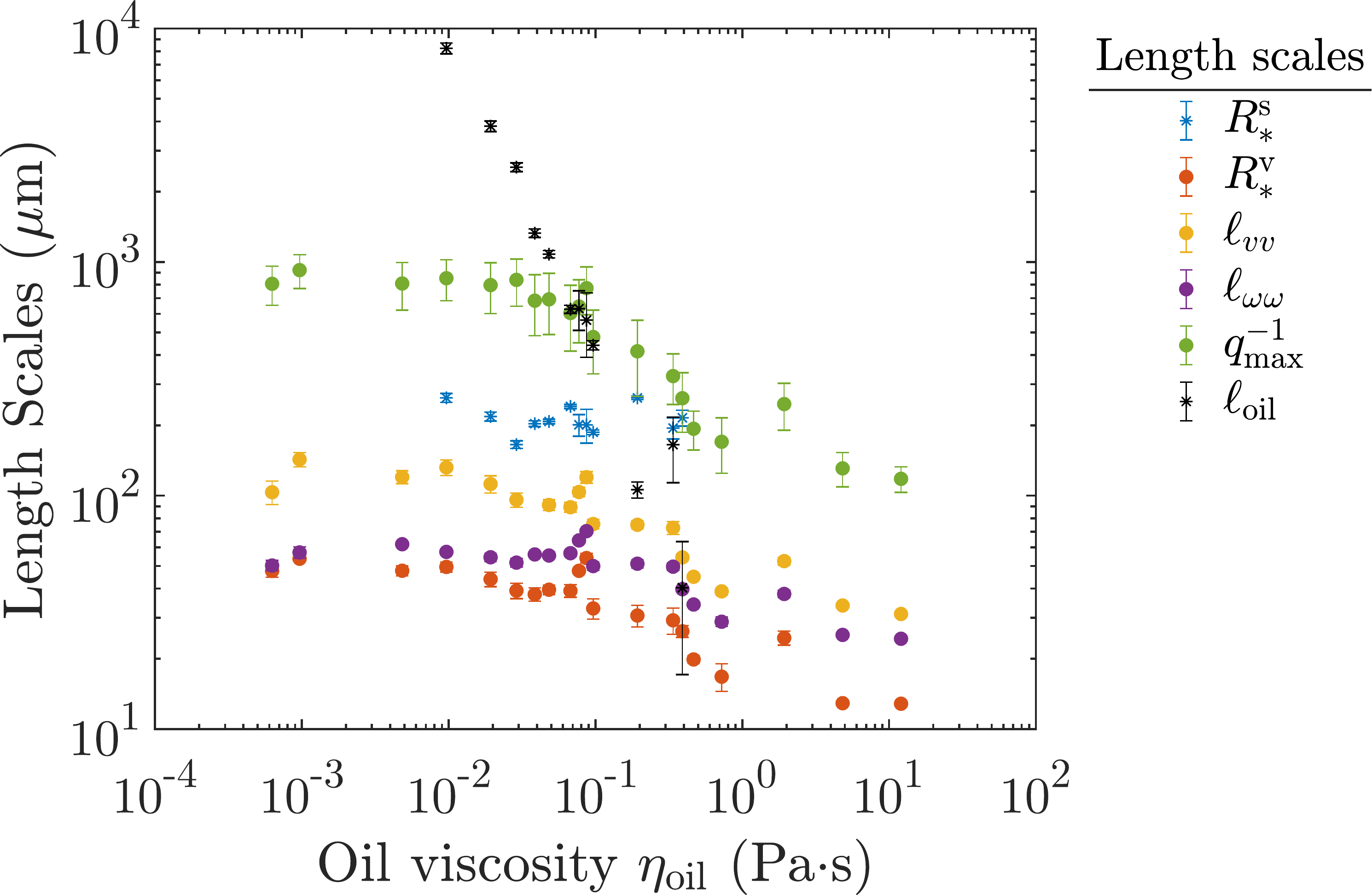}
\end{center}
\bfcaption{Comparison of characteristic length scales of active nematic turbulence}{ The different length scales are defined as follows: $R^{\text{s}}_*$ is the mean vortex radius obtained by fitting the kinetic energy spectra (\cref{Fig fit,Fig vortex-radius-spectra}); $R^{\text{v}}_*$ is the mean vortex radius obtained by fitting the exponential tail of the vortex area distribution (\cref{Fig vortex-areas,Fig vortex-radius-distribution}); $\ell_{vv}$ is the distance at which the velocity autocorrelation is 0.5 (\cref{Fig velocity-correlation}); $\ell_{\omega\omega}$ is the distance at which the vorticity autocorrelation is 0.5 (\cref{Fig vorticity-correlation}); $q_{\text{max}}^{-1}$ is the inverse of the wave number for which the kinetic energy spectrum has its maximum; $\ell_{\text{oil}} = \eta_{\text{n}}/\eta_{\text{oil}}$ is the oil viscous length obtained from the fits of the kinetic energy spectra (\cref{Fig fit}). Except for $\ell_{\text{oil}}$, all other length scales have similar behaviours with oil viscosity, consistent with them being proportional to one another.}
\label{Fig lengths}
\end{figure*}

\renewcommand{\figurename}{Movie}
\setcounter{figure}{0}
\begin{figure*}
	\begin{center}
		\includegraphics[width=1\linewidth]{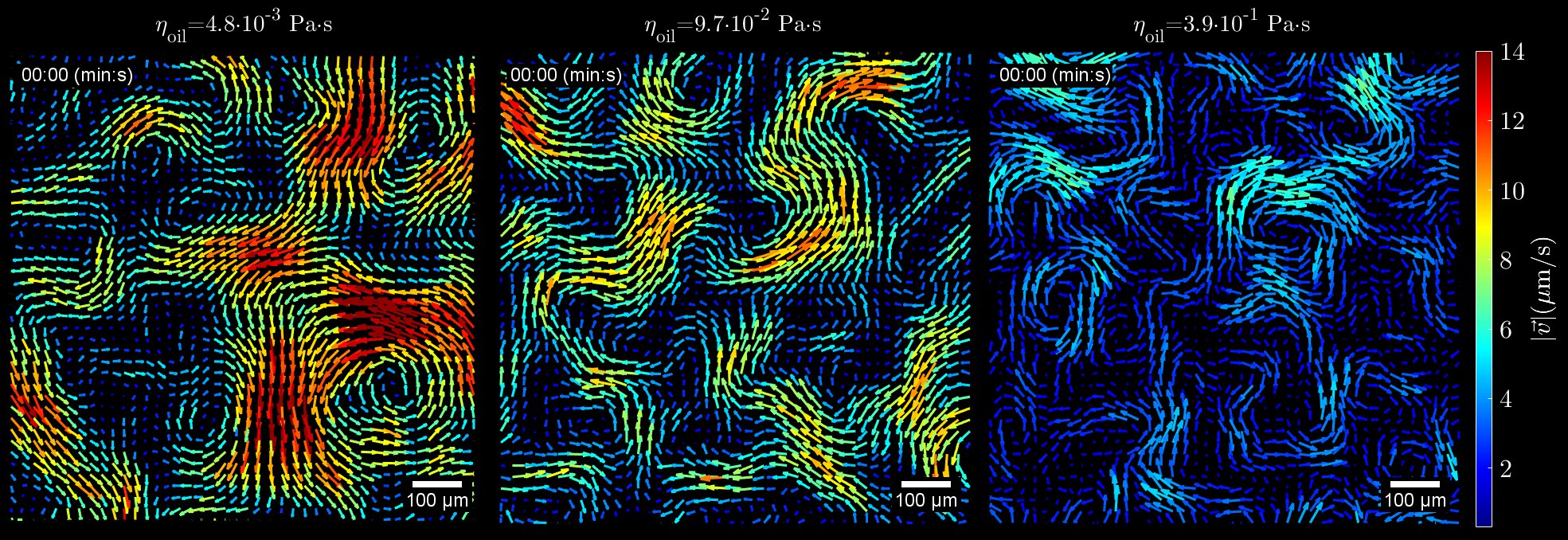}
	\end{center}
	\bfcaption{Flow field of the AN at different oil viscosities}{ As the oil viscosity is increased, the flows within the active layer become slower and the characteristic vortex size decreases. The velocity fields are the ones obtained from the PIV analysis of the experiments. Colors of the vectors indicate their magnitude. For the sake of a better visualization, we show a field of view smaller than the one used to compute the kinetic energy spectra and plot one vector every three. The movie is sped up x5.}
	\label{fig:movies1}
\end{figure*}

\end{document}